\documentclass[final,pre,twocolumn,graphicx,showpacs]{revtex4-2}
\usepackage{subfigure}
\usepackage{graphicx}
\usepackage{epsfig}
\usepackage{amsmath}
\usepackage{epstopdf} 
\usepackage{natbib}
\usepackage{color}
\usepackage{mathtools}
\usepackage{float}

\usepackage{hyphenat}
\usepackage[colorlinks=true,linkcolor=blue,urlcolor=blue,citecolor=blue]{hyperref}
\newcommand{\htilde}{{\tilde h}}
\newcommand{\etal}{{\it et al\ }}
\graphicspath{{./Figures/}}
\begin{document}


\title{Critical Behaviour of Magnetic Polymers in Two and Three Dimensions}
 
\author{Damien Paul Foster}\email{ab5651@coventry.ac.uk}
\affiliation{Centre for Computational Science and Mathematical Modelling,\\ Coventry University, Coventry CV1 5FB, UK}
\author{Debjyoti Majumdar}\email{debjyoti@iopb.res.in}
\affiliation{Institute of Physics, Bhubaneswar, Odisha 751005, India}
\affiliation{Homi Bhabha National Institute, Training School Complex, Anushakti
Nagar, Mumbai 400094, India}

\begin{abstract}
We explore the critical behaviour of two and three dimensional lattice models of polymers in dilute solution where the monomers carry a magnetic moment which interacts ferromagnetically with near-neighbour monomers. Specifically, the model explored consists of a self-avoiding walk on a square or cubic lattice with Ising spins on the visited sites. In three dimensions we confirm and extend previous numerical work, showing clearly the first-order character of both the magnetic transition and polymer collapse, which happen together. We present results for the first time in two dimensions, where the transition is seen to be continuous. Finite-size scaling is used to extract estimates for the critical exponents and transition temperature in the absence of an external magnetic field.  
\end{abstract}

\maketitle

\section{Introduction}

Self-avoiding walk models on lattices have been used for several decades as good models for polymers in solution\cite{vanderzande_1998}. Short-ranged interactions between nearest-neighbour visited lattice sites are introduced to mimic the effects of solvent quality, and other interactions may be introduced to account for other effects.  Canonically, if the interactions are short-ranged and the walks studied in their infinite length limit, there should exist a universality where the critical behaviour is insensitive to details of the model\cite{de_gennes_collapse_1975,Stephen1975,Duplantier1982}. An early indication that things are not so simple arose in the study of the two-dimensional $O(n\to 0)$ model introduced by Bl\"ote and Nienhuis\cite{Blote1989}, more recently known under the title of Vertex Interacting Self-Avoiding Walk~\cite{foster_corner_2003,foster_surface_2012,bedini_numerical_2013,Foster2019}. They found that in this model, despite having short-ranged interactions, corresponding to non-crossing doubly-visited sites, the critical exponents were not the same as those found previously for the interacting self-avoiding walk (ISAW). This lack of universality is due to a fractal dimension for the walk which equals that of the lattice, giving rise to a critical point with a first-order character\cite{foster_corner_2003,Vernier2015}. 

\begin{figure}[H]
\includegraphics[width=8cm]{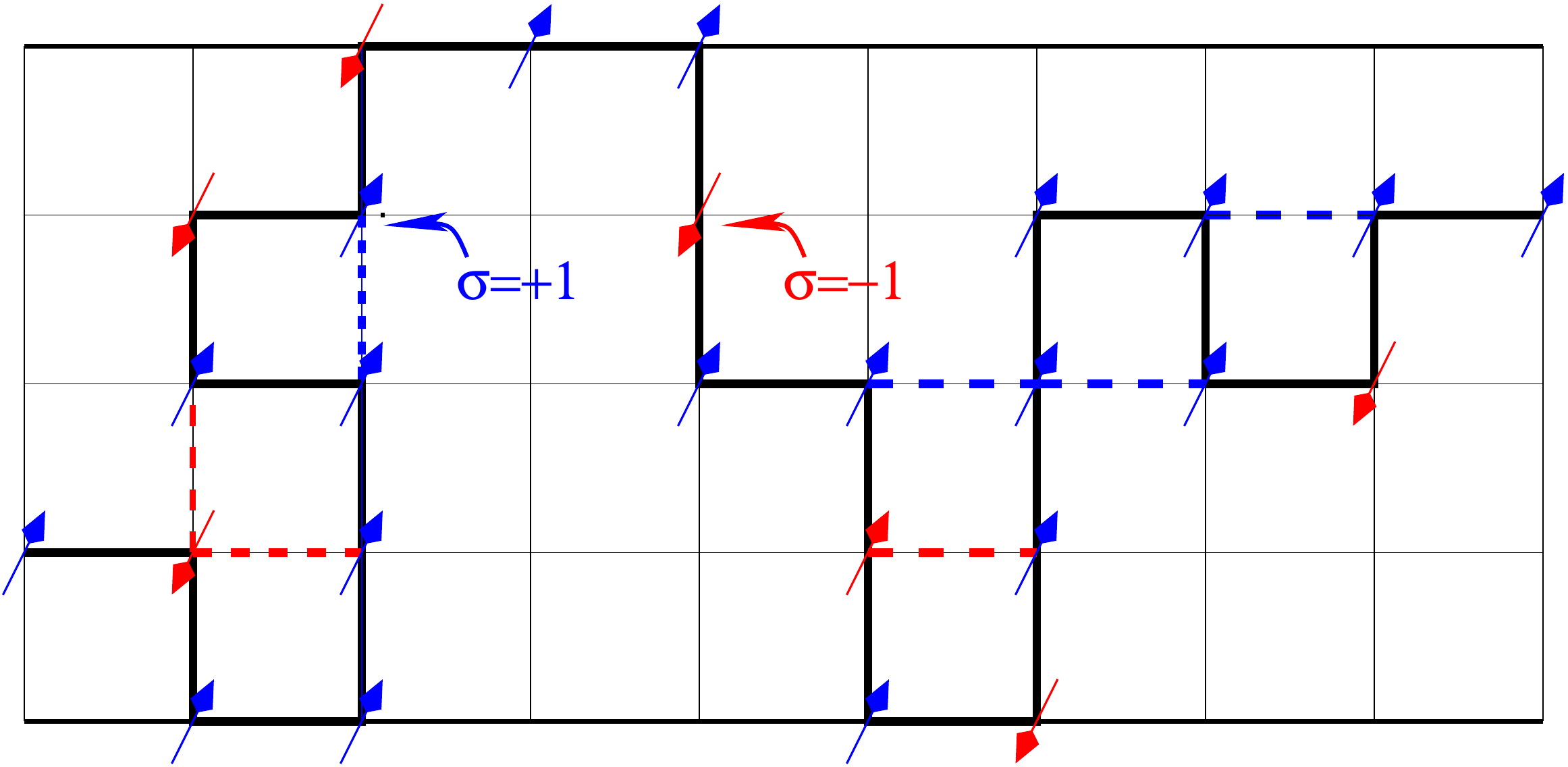}
\caption{magnetic walk model in two dimensions}\label{model}
\end{figure}

In the models described above, the walks are neutral and non-magnetic. In this paper we look at the critical behaviour of polymers where each monomer has a magnetic moment (spin) which may be either ``up" or ``down". The walk is modelled by a self-avoiding walk on a square (2d) or cubic (3d) lattice and the spins sit on the occupied lattice sites and interact via the standard Ferromagnetic Ising Hamiltonian (Figure~\ref{model}). 
The spins interact with all spins that are on adjacent sites (including along the walk). The only energy taken into account is this Ising interaction energy, however the entropy will be the sum of the spin entropy and the walk configurational entropy.  In three dimensions, this model was introduced by Garel {\it et al}\cite{Garel1999} and studied in three dimensions.

The partition function for a walk of length $N$ is
\begin{equation}
Z_N=\sum_{\rm \Omega_N}\sum_{\{\sigma_i=\pm 1\}}\exp\left(\beta J\sum_{\langle i,j \rangle} \sigma_i \sigma_j+\beta h \sum_i \sigma_i\right),
\end{equation}
where $\Omega_N$ is the set of self-avoiding walks of length $N$, \mbox{$\sigma_i=\pm 1$} are the two-state Ising spins, and as usual $\langle i,j\rangle$ indicates that the sum is over pairs of spins which are nearest-neighbour on the lattice, $h$ is a magnetic field and \mbox{$\beta=1/kT$} as usual. In what follows we will set $J/k=1$.

It is important to be clear that equilibrium is defined by a minimisation of the free energy 
\begin{equation}
F=E_{\rm Ising}-T(S_{\rm Ising}+S_{\rm SAW}),
\end{equation}
In other words, the polymer configurations and the spin states are both fluctuating quantities in thermal equilibrium.
There have been many studies where either one or the other are frozen, i.e. the spin configuration is quenched and the walk configurations studied\cite{Archontis1994} or the self-avoiding walk configuration is quenched, and the behaviour of the Ising model on the resulting fractal studied\cite{Chakrabarti1983,Bhattacharya1984,Chakrabarti1985}, but the current situation has been much less studied. 
The model described here was introduced by Garel, Orland and Orlandini\cite{Garel1999} and studied by both Mean-Field Theory (MFT) and 
Parallel Tempering Monte-Carlo method in 3d. The MFT suggested that at low magnetic fields there is a simultaneous first-order magnetic and collapse transition. In the limit of infinite magnetic field, the model is simply the usual ISAW model, where the transition is tricritical\cite{de_gennes_collapse_1975}, and this extends into a line of tricritical transitions as $h$ is lowered. The first-order and tricritical lines are separated by a multicritical point. Their numerical results tend to support the MFT picture. The main evidence for the behaviour at $h=0$ was a rapid variation of the magnetisation and radius of gyration as the transition was approached, and linear scaling of the specific heat as an indicator of the first order transition.

Another related model was studied by Luo and coworkers~\cite{Luo2003,Luo2003a,Huang2004,Huang2005,Luo2006,Luo2006a} in zero external magnetic field in 3d on the cubic lattice. In this case they allowed the bonds to fluctuate in length, but pairs of spins only interacted if they were nearest-neighbours, either along the chain, or not. Luo concluded that the transition was critical, and gave an estimate of the critical temperature and some exponents\cite{Luo2006}. 

In this paper we revisit the three dimensional model using flatPERM method\cite{prellberg_flat_2004} to stochastically enumerate the number of configurations as a function of the number of magnetic energy indexed by $n_i$ below, and the magnetisation indexed by $n_s$ from which we can construct the 
partition function, given by
\begin{equation}
Z_N=\sum_{n_i=-n_{i, \rm min}}^{n_{i, \rm max}} \sum_{n_{s}=-N-1}^{N+1} C_{N,n_i,n_{s}} \exp\left(\frac{1}{T}(n_i+ h n_{s})\right),
\end{equation}
where $n_i = \sum_{\langle i,j\rangle} \sigma_i\sigma_j$ is the difference between the number of satisfied nearest-neighbour bonds and the number of unsatisfied bonds and $n_s=\sum_{i=0}^N \sigma_i$  
the difference between the number of up spins and the number of down spins.  The internal energy $\langle E \rangle$ and the square magnetisation $\langle M^2 \rangle$ can be directly calculated (the magnetisation is expected to be zero on average at all temperatures by symmetry). As we will also be interested in the fourth order Binder cumulant\cite{Binder2010}, we will also need to calculate $\langle M^4\rangle$. These averages can be calculated in the usual way using the expressions
\begin{eqnarray}
\langle E^x \rangle &=& \frac{1}{Z_N} \sum_{n_i, n_{s}}  n_i^x C_{N,n_i,n_{s}} \exp\left(\frac{1}{T}(n_i+h n_s)\right),\\
\langle M^x \rangle &=& \frac{1}{Z_N} \sum_{n_i, n_{s}} n_{s}^x C_{N,n_i,n_{s}} \exp\left(\frac{1}{T}(n_i+ h n_{s})\right).
\end{eqnarray}

The fourth-order Binder cumulant is defined as\cite{Binder1981,Binder1981a} 
\begin{equation}
U_m=1-\frac{\langle M^4 \rangle}{3\langle M^2\rangle^2}.
\end{equation}
As the number of spins tends to infinity, $U_m\to 0$ from above for $T>Tc$ and $U_m\to 2/3$ from below for $T<T_c$, at $T_c$, $U_m\to U^*$, which is a unique number related to geometry and boundary conditions, and so not universal\cite{Chen2004,Selke2006}. Crossings of the Binder cumulant are indicative of a phase transition\cite{Binder1981}.

A phase transition is also indicated, typically, by a divergence of the specific heat $C=\frac{1}{NT^2}\left(\langle E^2 \rangle - \langle E \rangle^2\right)$. Finite-size estimates of the transition temperature can be obtained from the peak of these fluctuations.  

Any other quantity can be calculated if the table of values is accumulated as the configurations are enumerated. Here we have calculated the radius of gyration squared 
\begin{equation}
R_g^2=\frac{1}{Z_N} \sum_{n_i, n_s}  r_{N,n_i,n_s} \exp\left(\frac{1}{T}(n_i+h n_s)\right),
\end{equation}
where
\begin{equation}
r_{N,n_i,n_s}=\sum_{i\in\Omega_{N,n_i,n_s}}  \left\langle \vec{r_i}-\langle \vec{r_i} \rangle\right\rangle^2
\end{equation}
and $i$ labels the chain in the set $\Omega_{N, n_i, n_s}$ of walks of length $N$ with $n_i$ and $n_s$ fixed. The average is done over the positions of the $N+1$ occupied sites of chain $i$. 

In order to look at longer chain lengths, we can define partial partition functions for fixed $n_i$, but incorporating the magnetic field:
\begin{equation}
C^{\rm eff}_{N, n_i}=\sum_{n_s=-N+1}^{N+1}  C_{N,n_i,n_s} \exp\left(-\frac{h}{T} n_s\right).
\end{equation}
Everything proceeds as before, but $M^2$, $M^4$ and $|M|$ need to be accumulated in the same way as $R_g^2$ to enable weighted averages to be calculated later.

\section{Scaling relations}

In polymer physics, the exponents have slightly different interpretations and sometimes expressions than for the Ising model, and in this model there is a risk of confusion between the two.  In this section we take the opportunity to recall the relevant finite-scaling relations which will be used to identify the transition points and the critical exponent estimates, since we are working in the fixed length ensemble, and not the usual fixed lattice size ensemble. 

If we cast the problem on an infinite lattice, and control the length of the walk through a fugacity $K$, the partition function would be given by:
\begin{equation}
{\cal Z}=\sum_{N=0}^{\infty} K^N Z_N.
\end{equation}
The (average) length of the walk is governed by $K$, and $\langle N \rangle\sim (K_c-K)^{-1}$ for $K\leq K_c$. The free energy per lattice site for $K\leq K_c$ is 0, which is a reflection of the fractal nature of the walk. As a result one uses the free-energy per monomer instead, and the correct scaling expression for the singular part of the free energy per monomer (and hence spin) is 
\begin{equation}\label{free_en}
f_s=b^{-d_H}\tilde f(k b^{y_1},t b^{y_2},h b^{y_h}),
\end{equation}
where $b$ is the linear scaling factor, where the Hausdorff dimension of the walk $d_H\leq d$. The $\{y_i\}$ are the scaling dimensions\cite{cardy1996}. We have defined reduced dimensionless variables
\mbox{$k=(K_c-K)/K_c$} and $t=(T_c-T)/T_c$. Typically, we identify $d_H=y_1=1/\nu$, where $\nu$ is the geometric exponent given by 
\begin{equation}\label{rscaling}
R\sim N^\nu,
\end{equation}
where $R$ is any typical linear dimension of the walk, eg. radius of gyration, end-to-end distance or hydrodynamic radius. 

Taking two derivatives with respect to $t$, and then setting $tb^{y_2}$ constant gives:
\begin{equation}
C\sim t^{y_1/y_2-2}\sim t^{-\alpha},
\end{equation}
where we have used $d_H=y_1$ and $y_2/y_1=\phi$, leading to the definition of  $\alpha=2-1/\phi$, relevant for polymer models\cite{lam_specific_1987}.

In order to introduce the finite-size scaling relations we need, we identify the relevant scale factor as the radius of gyration $r=\sqrt{\langle R_g^2\rangle}$, or alternatively $N^\nu=N^{1/y_1}$. Fixing $N$ corresponds to fixing $k$, and using~\ref{free_en}, we find the tricritical scaling expression:
\begin{equation}
f_s=N^{-1}\tilde f(kN,tN^\phi,hN^{\Delta})=N^{-1}\Phi(tN^\phi,hN^\Delta),
\end{equation}
where $\Delta=y_h/y_1$.

Fixing \mbox{$tN^\phi=x_{\rm max}$}, the value giving the maximum of $C$, and differentiating twice $f_s$ again with respect to $t$, gives
\begin{eqnarray}
C_{\rm max}&\sim& N^{-1+2\phi}\sim N^{\alpha\phi}\sim N^{\frac{\alpha}{2-\alpha}}\ {\rm and}\\
t&\sim& N^{-\phi},
\end{eqnarray}
for sufficiently large $N$.

For a first-order transition, we expect $C_{\rm max}$ to scale with the volume, which here is $N$, leading to $\alpha=1$ at the first-order transition expected in three dimensions and $h=0$. This is what was observed by Garel {\it et al.}\cite{Garel1999}, and is confirmed here. At the standard collapse transition (relevant for larger $h$), $\alpha=0$ with logarithmic corrections\cite{Duplantier1982,Duplantier1987}.

We will also use the scaling behaviours of \mbox{$m=\frac{1}{N+1}\sqrt{\langle M^2 \rangle}$} and \mbox{$r=\sqrt{\langle R_g^2\rangle}$}.

For the magnetisation, we expect either \begin{equation} m\sim (T_c-T)^{\beta}\end{equation} if the transition is critical, or a jump in $m$ if first-order.
The usual finite-size scaling for the ferromagnetic transition would be
\begin{equation}
\frac{1}{(N+1)^2}\langle M^2 \rangle=L^{-2\beta/\nu_2}\tilde M(tL^{y_2}).
\end{equation}
 The relevant length scale is again $L=r=aN^\nu$, and so:
\begin{equation}\label{magscaling}
m=N^{-\beta\phi}\tilde m(tN^{\phi}).
\end{equation}
%

Assuming the scaling behaviour for $r$ and $m$ from Equations~\ref{rscaling} and~\ref{magscaling}, we can define scaling functions
\begin{eqnarray}\label{nightingale}
\varphi_{R_g}&=&\frac{\log\left(r_N/r_{N^\prime}\right)}{\log\left(N/N^{\prime}\right)},\\\nonumber
\varphi_m&=&\frac{\log\left(m_N/m_{N^\prime}\right)}{\log\left(N/N^{\prime}\right)}.
\end{eqnarray}
Crossings of these functions give estimates of the $T_c$\cite{Nightingale1976}, which are expected to converge to the correct value as $N\to \infty$. Also, as $N\to\infty$, the values of these functions give \mbox{$\varphi_{R_g}\to \nu$} and \mbox{$\varphi_m\to -\beta\phi$}.  

Another useful tool to use for estimating critical temperatures and, in the case of polymers, the exponent $\phi$ is to look at the behaviour of the dominant complex zero of the partition function\cite{lee_exact_2010,ponmurugan2012,taylor_partition_2014,Foster2019}. The dominant zero is the one which is closest to the real axis in the complex temperature plane, and will pinch the real axis at the transition temperature in the thermodynamic limit. It is found that the real and imaginary parts of the zero behave, to leading order, as
\begin{eqnarray}
T_r&\sim& T_c+a N^{-\phi}\\  
T_i&\sim& b N^{-\phi}.
\end{eqnarray}

In practice, we will calculate the zeros in terms of the relevant Boltzmann weights, for example $\tau=\exp(-1/T)$, as the flatPERM method gives us a polynomial in $\tau$. The complex zeros of this polynomial are then calculated using the MPsolve package\cite{BINI2014276,2000NuAlg..23..127B}.
 
 \section{Results}

One of the advantages of the flatPERM stochastic enumeration method is that you calculate the coefficients of the series expansion in terms of the dependent variables $1/T$ and $\htilde=h/T$. This means that you can plot quantities of interest as functions of these parameters, and not simply isolated points as in traditional Monte-Carlo methods, including those used by Garel \etal in Ref\cite{Garel1999}. There is a price to be paid, however, which is that the more dependent variables you keep track of, the slower the method is to converge, and the shorter the maximal chain length attainable in practical times. It is therefore useful to take cuts at fixed values of $\htilde$, to be able to explore longer chains. Here we do both. The longest chains we consider are of length $N=600$ in 3d and $N=1000$ in 2d. Typically we looked at ten independent runs of about $10^7$ tours each. The ten different runs were used to ensure that the results had converged, and the final results used a series that was an average over the 10 sets.

 \subsection{Three-dimensional magnetic self-avoiding walks}

\begin{figure}[H]
\includegraphics[width=8cm]{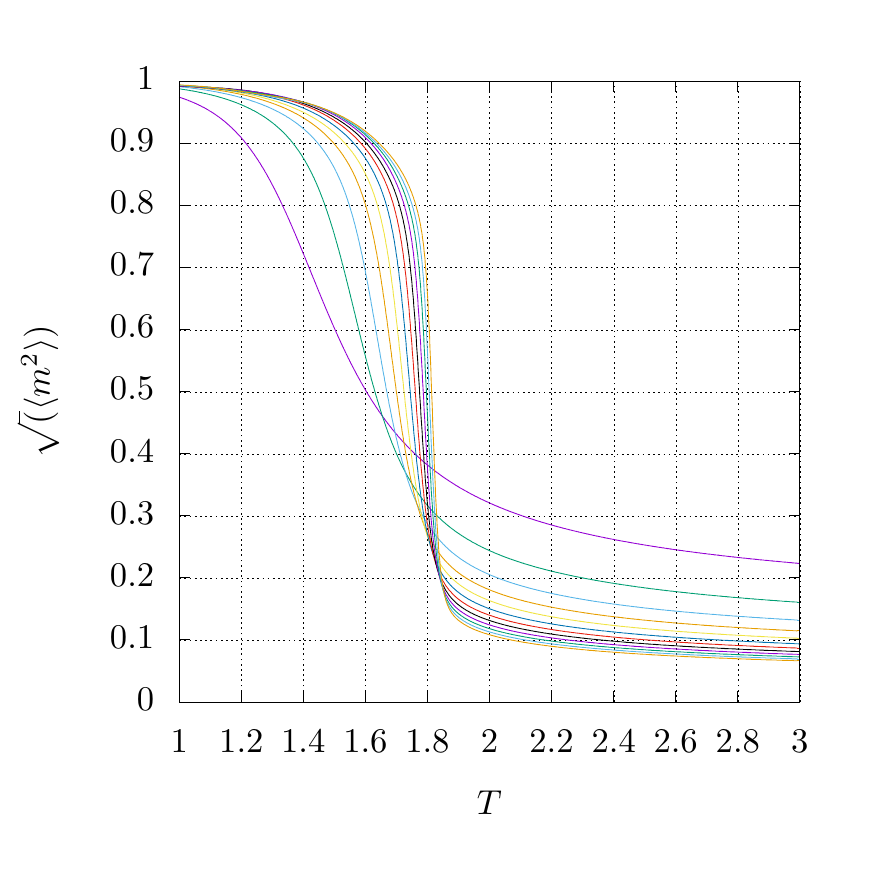}
\caption{Magnetisation as a function of $T$ in 3d for $h=0$ shown for sizes from $N=50$ to $N=600$ in steps of 50.}\label{mag3dh0}
\end{figure}

\begin{figure}[H]
\includegraphics[width=8cm]{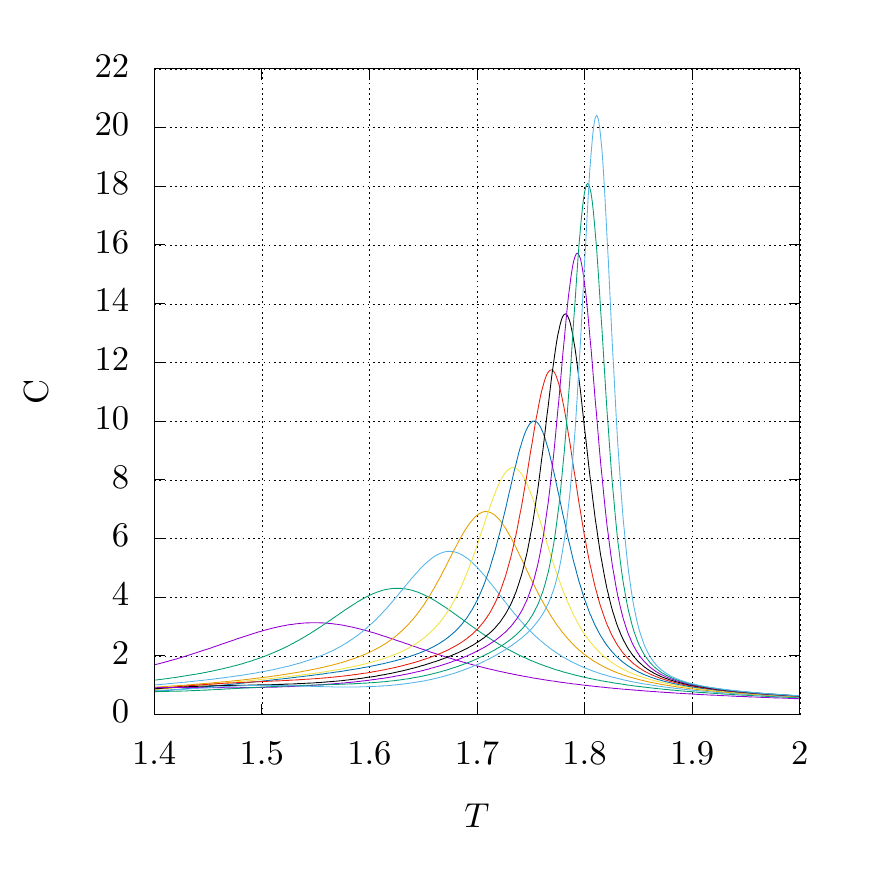}
\caption{Specific heat $C$ as a function of the temperature for sizes $N=100$ to $N=600$ in steps of 50}.\label{3dres}
\label{3dh0}
\end{figure}

In Figures~\ref{mag3dh0}--\ref{3dres} we plot the magnetisation, radius of gyration per monomer (proportional to the inverse of the density), the Binder cumulant and the specific heat as a function of $T$. The Binder cumulant shows a characteristic negative spike, indicative of a first-order transition\cite{Binder2010}. 
%


\begin{figure}[H]
\includegraphics[width=8cm]{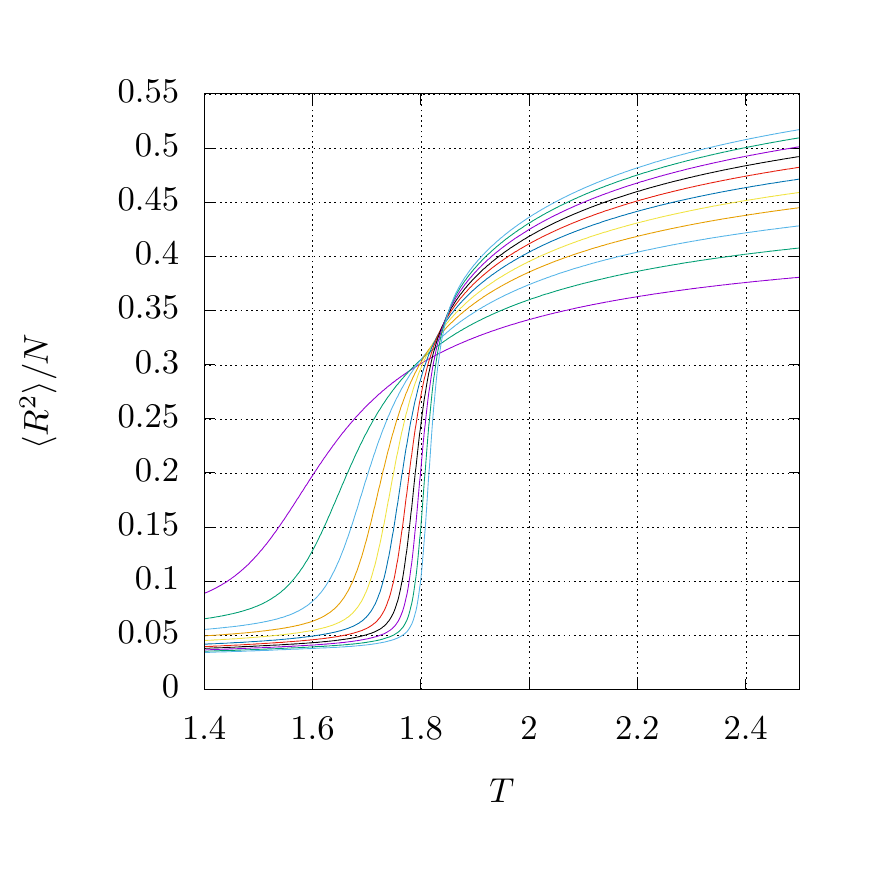}
\caption{Radius of gyration as a function of $T$ in 3d for $N=50$ to $N=600$ in steps of 50. }
\end{figure}

\begin{figure}[H]
\includegraphics[width=8cm]{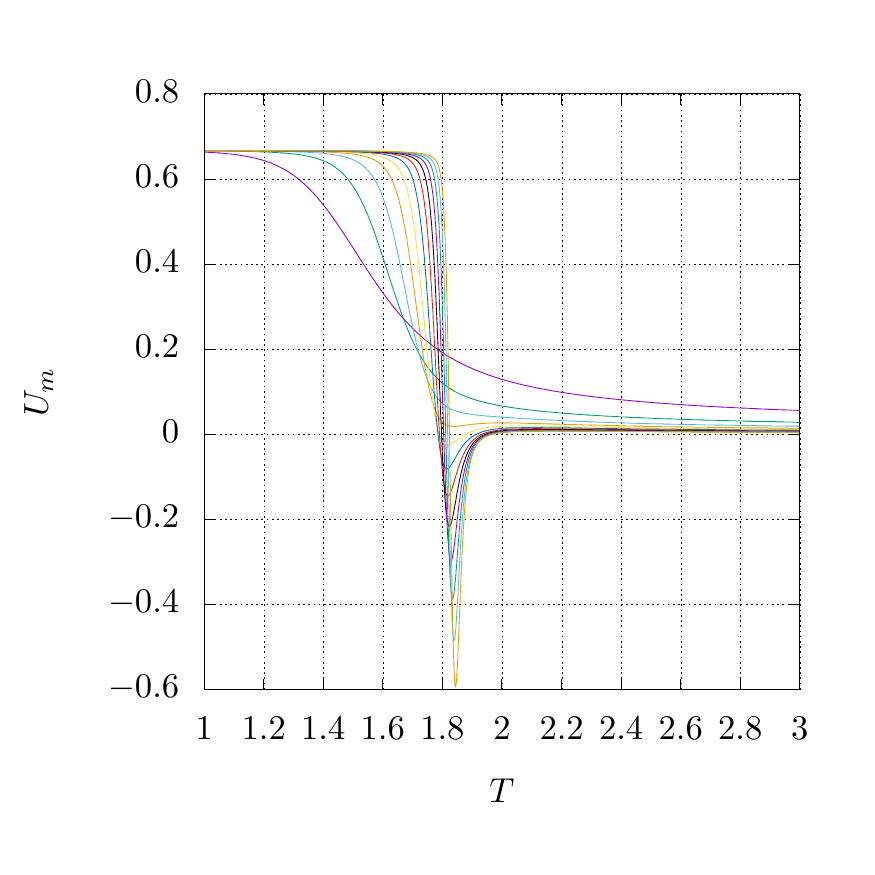}
\caption{Binder cumulant plotted for $h=0$ in 3d for $N=50$ to $N=600$ in steps of 50.}
\end{figure}

\begin{figure}[H]
\includegraphics[width=8cm]{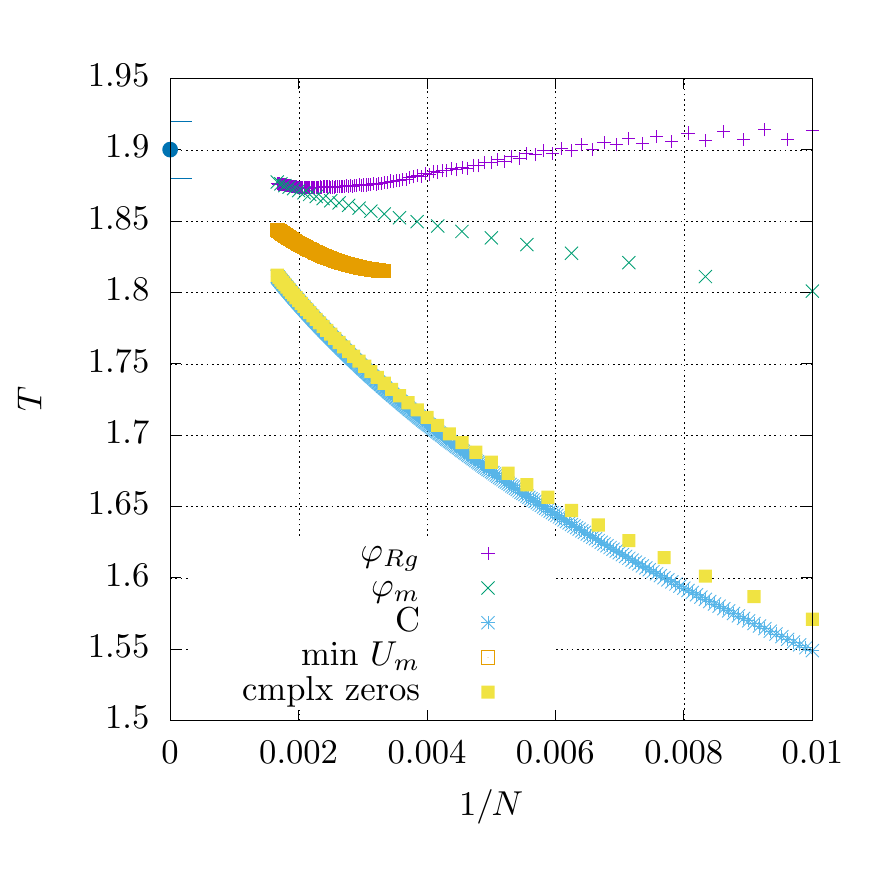}
\caption{Finite-size estimates of the critical temperature derived from crossings of $\varphi_{R_g}$, $\varphi_m$, $C_{\rm max}$ and the minimum of $U_m$, as well as the locus of the dominant zero of the partition function as a function of $1/N$.}\label{tc3dh0}
\end{figure}

\begin{figure}
\includegraphics[width=8cm]{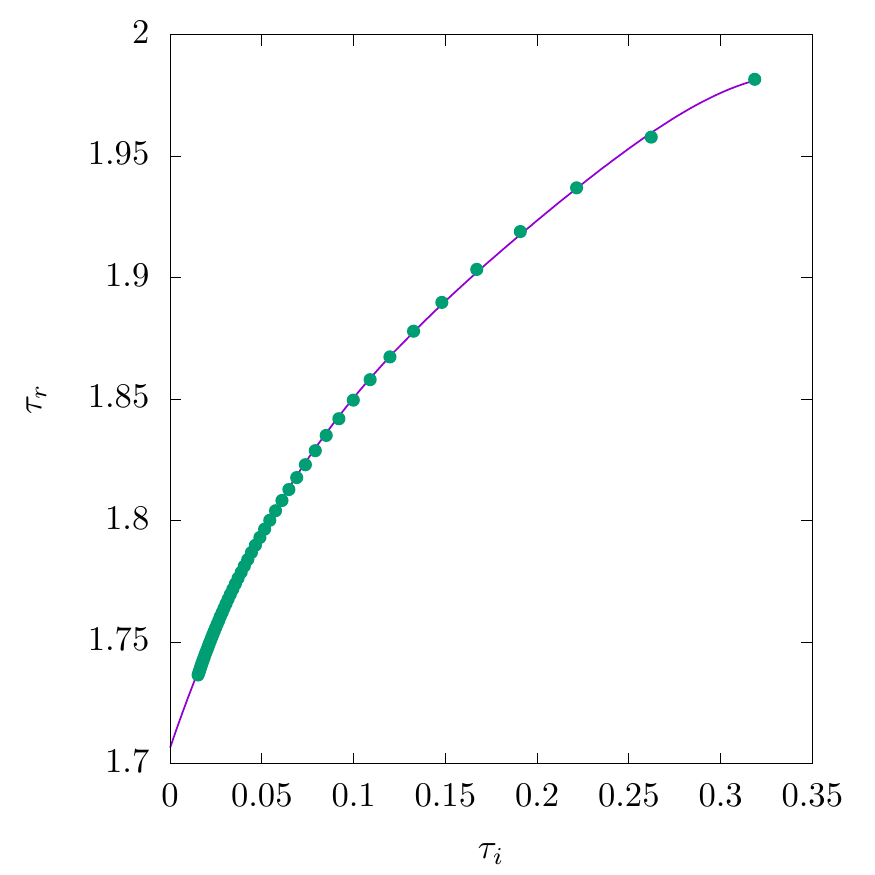}
\caption{Estimates of the critical temperature for $h=0$ by extrapolation of $\tau_r$ as $\tau_i\to 0$ where $\tau=\exp(-1/T)$. Points fitted by a quartic equation in $\tau_i$.
$\tau_c=1.7067$ or $T_c=1.871$.}
\label{cplx3d}
\end{figure}

We calculated different finite-size estimates of the transition temperature using different methods: Nightingale phenomenological renormalisation group\cite{Nightingale1976} with the scaling relations~\ref{nightingale}, looking at the maximum of the specific heat and the real part of the leading complex partition function zero. We also used the minimum of $U_m$ as an indicator of the transition temperature (the bottom of the spike). These estimates are shown in Figure~\ref{tc3dh0}. The different estimates converge to give an estimate of $T_c=1.90\pm 0.02$. As an alternative method, we also looked at the locus of the leading complex zero in the complex plane\cite{Lee1952,Yang1952,Fisher1965}. In Figure~\ref{cplx3d} we plot the zeros in terms of the Boltzmann weight $\tau=\exp(-1/T)$. Expanding $\tau_r$ as a function of powers of $N^{-\phi}$, and using $\tau_i\propto N^{-\phi}$, we can look at fitting $\tau_r$ as a polynomial in $\tau_i$ to extrapolate to $\tau_i\to 0$. In Figure~\ref{cplx3d} we used a quartic equation. This extrapolation gave an estimate of $\tau_c=1.7067$, or a transition temperature of $T_c=1.871$.

\begin{figure}[H]
\includegraphics[width=8cm]{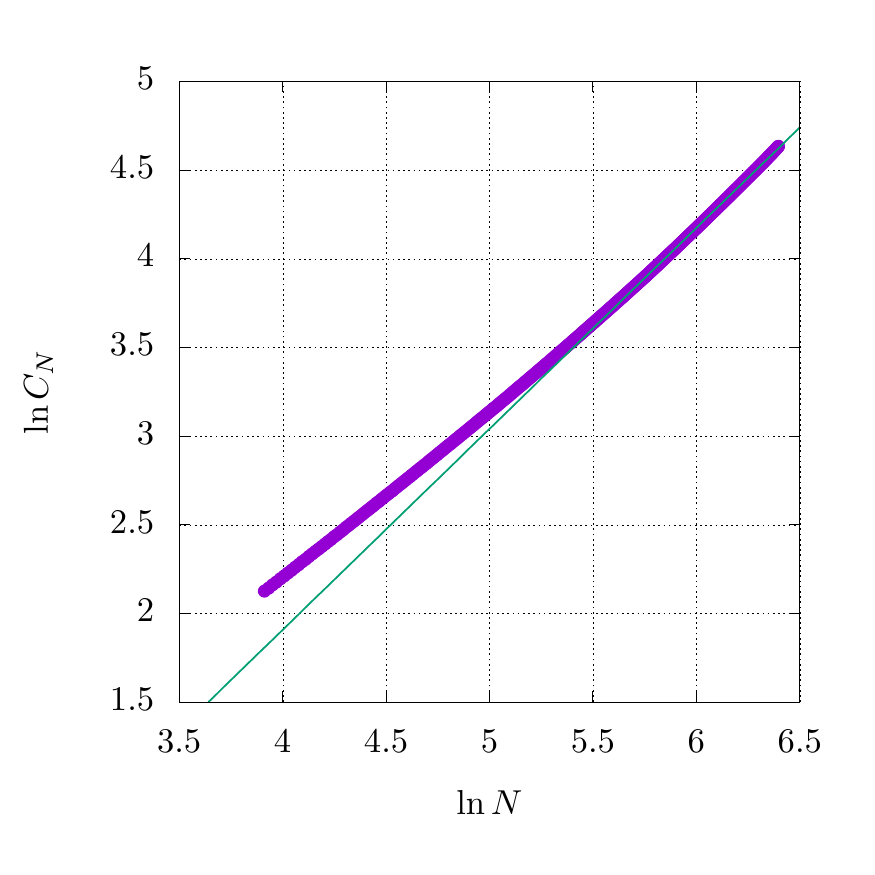}
\caption{Plot of $\ln C_{\rm max}$ vs $\ln N$. The guide line has a slope of 1.13, which translates to an exponent $\alpha=1.06$, which is only slightly larger than 1, expected for a first-order transition.}\label{logC}
\end{figure}

\begin{figure}[H]
\includegraphics[width=8cm]{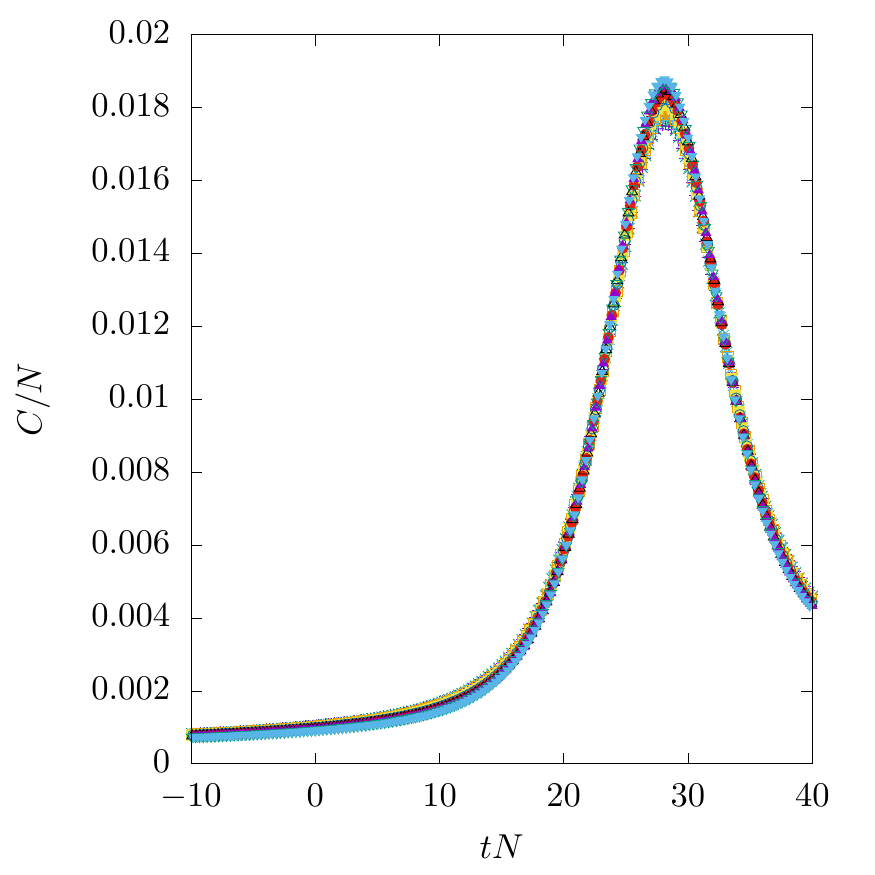}
\caption{Scaled specific heat $C/N$ plotted against the scaled reduced temperature $tN$ with $T_c=1.9$ and sizes from $N=500$ to $N=600$ in steps of 10.}\label{scaledC}
\end{figure}

\begin{figure}[H]
\includegraphics[width=8cm]{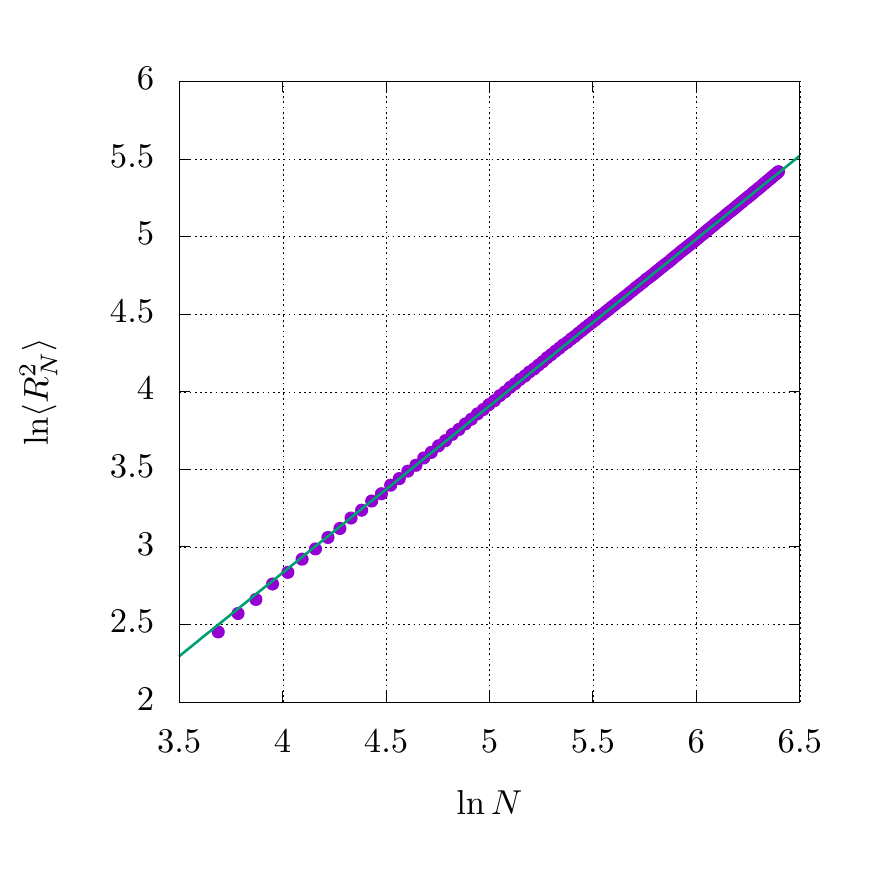}
\caption{The log-log plot of $\langle R_g^2 \rangle$ against $N$. The slope of the line, for comparison, is 1.075.}\label{logrg}
\end{figure}
\begin{figure}[H]
\includegraphics[width=8cm]{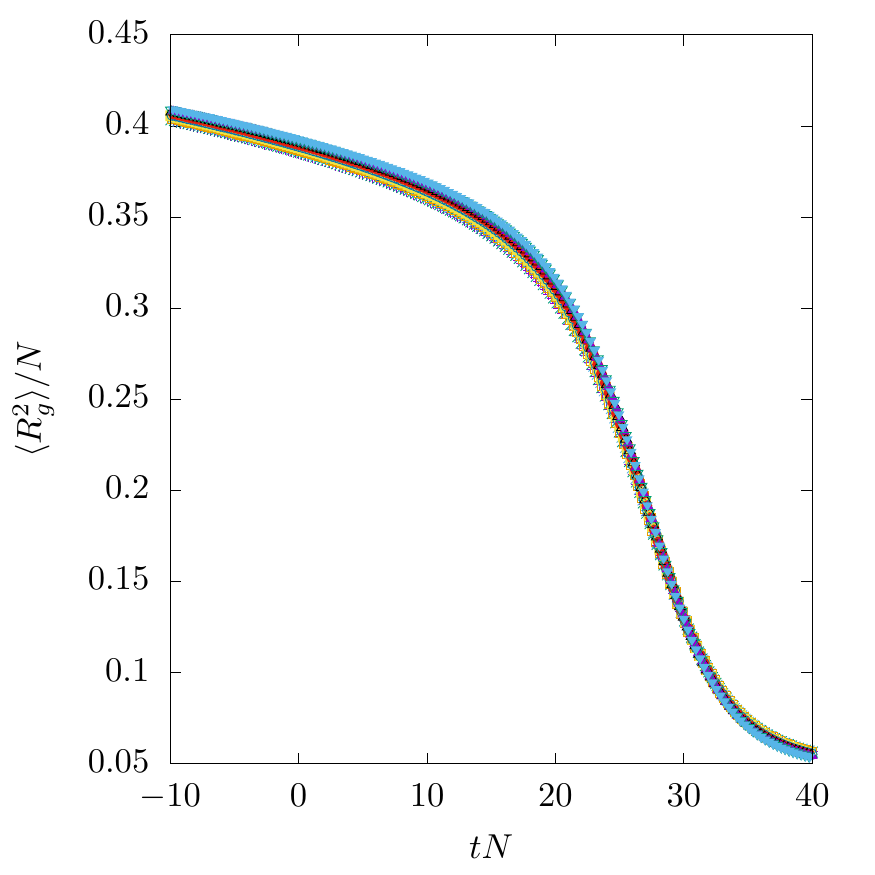}
\caption{scaled radius of gyration $\langle R_g^2\rangle/N$ against scaled reduced temperature $tN$with $T_c=1.9$ and sizes from $N=500$ to 600 in steps of 10.}\label{scaledRG}
\end{figure}

\begin{figure}[H]
\includegraphics[width=8cm]{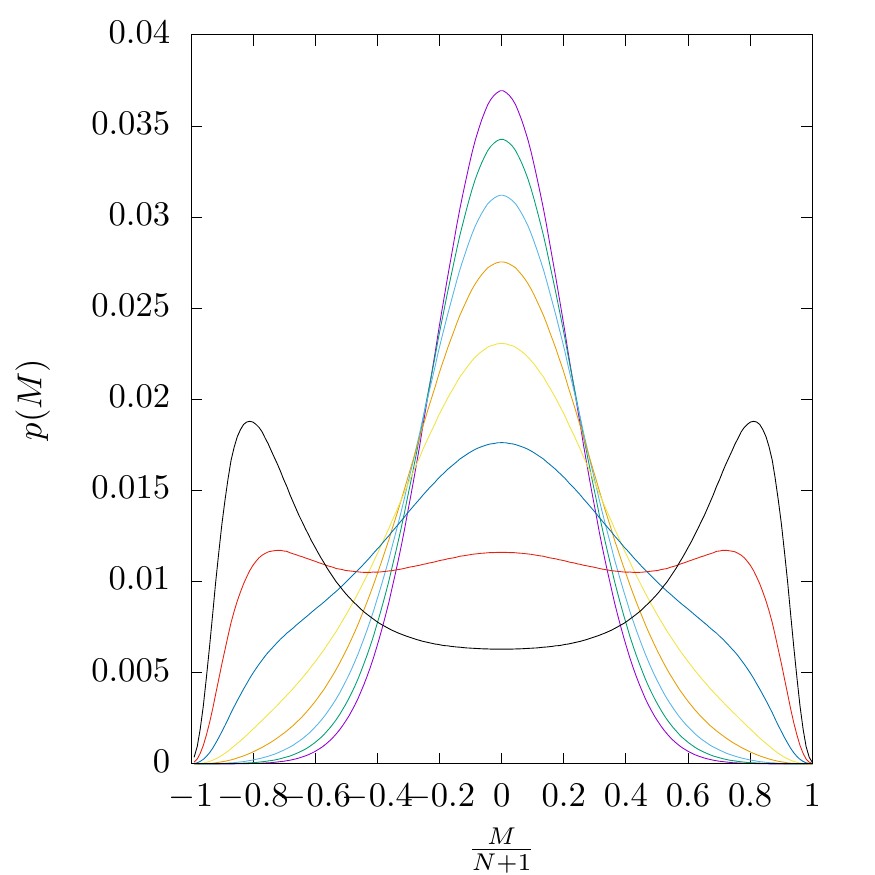}
\caption{Probability distributions of $M$ for $\beta$ from $0.5$ to $0.6$ in steps of $0.01$ for $N=200$.}\label{probM3d}
\end{figure}
\begin{figure}[H]
\includegraphics[width=8cm]{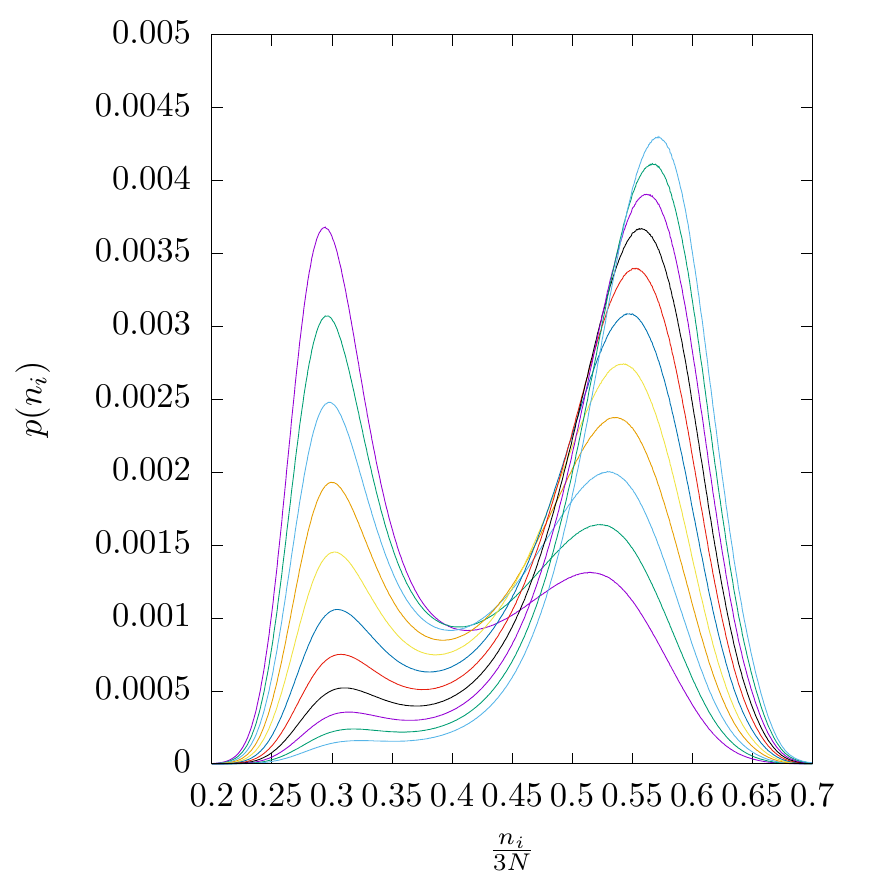}
\caption{Probability distributions of $n_i$ for $\beta$ from 0.4 to 0.55 in steps of 0.01 for $h=0$ for $N=600$.}\label{probRg3d}
\end{figure}



In Figures~\ref{logC} and~\ref{logrg} we plot log-log graphs of $C_{\rm max}$ and $r^2$ against $N$. The slope should give the exponent $\alpha/(2-\alpha)$ and $2\nu$ respectively. The $C_{\rm max}$ curve gives a slope corresponding to $\alpha/(2-\alpha)=1.13$, or $\alpha=1.06$ which is close to the expected linear behaviour. 


To confirm these exponents, we have defined the reduced temperature as \[t=\frac{T_c-T}{T_c}\] and plotted $C_N/N$ against $tN$ and $r^2/N$ against $tN$. The best collapse at $t=0$ was achieved  with $T_c=1.9$. These curves are shown in Figures~\ref{scaledC} and~\ref{scaledRG}. It is curious that the scaling of $C$ and $U_m$ indicate clearly a first-order transition, whilst the scaling radius of gyration seems to indicate $\nu=1/2$, as with the usual tricritical collapse. To investigate this further, we looked at the the probability distributions for different temperatures (or $\beta=1/T$). In order to calculate the probability distribution for $M$ (Figure~\ref{probM3d}) we need the full histogram in $n_i$ and $n_s$, and were limited to $200$, whilst we also looked at the probability distribution for the energy (expressed as $n_i/(3N)$) for $N=600$. For the magnetisation, we can clearly see the coexistence of three phases at $\beta=0.59$ or $T=1.69$, and in the energy distribution we can see clearly the coexistence of two phases, one with a small number of contacts and one with a larger number of contacts, which we interpret as swollen and collapsed phases. This latter explains the three peaks in the magnetisation: when the walk is extended, the Ising model is essentially one dimensional, and cannot order, giving a zero average magnetisation, whilst when the walk collapses, the support is three dimensional, and the Ising model is already in the magnetised phase. The tip-over point in energy occurs for $\beta$ between 0.52 and 0.53 ($T=1.92$ and $T=1.88$) for $N=600$. Note that the lower temperature for the magnetisation is mainly because of the smaller size of walk considered. It remains interesting to note, that whilst the collapse transition as well as the magnetic transition occur together, the average radius of gyration, taken across the two co-existing phases, still gives an exponent $\nu=1/2$ for the radius of gyration.


\subsection{Two-dimensional magnetic self-avoiding walks}

\begin{figure}[H]
\includegraphics[width=8cm]{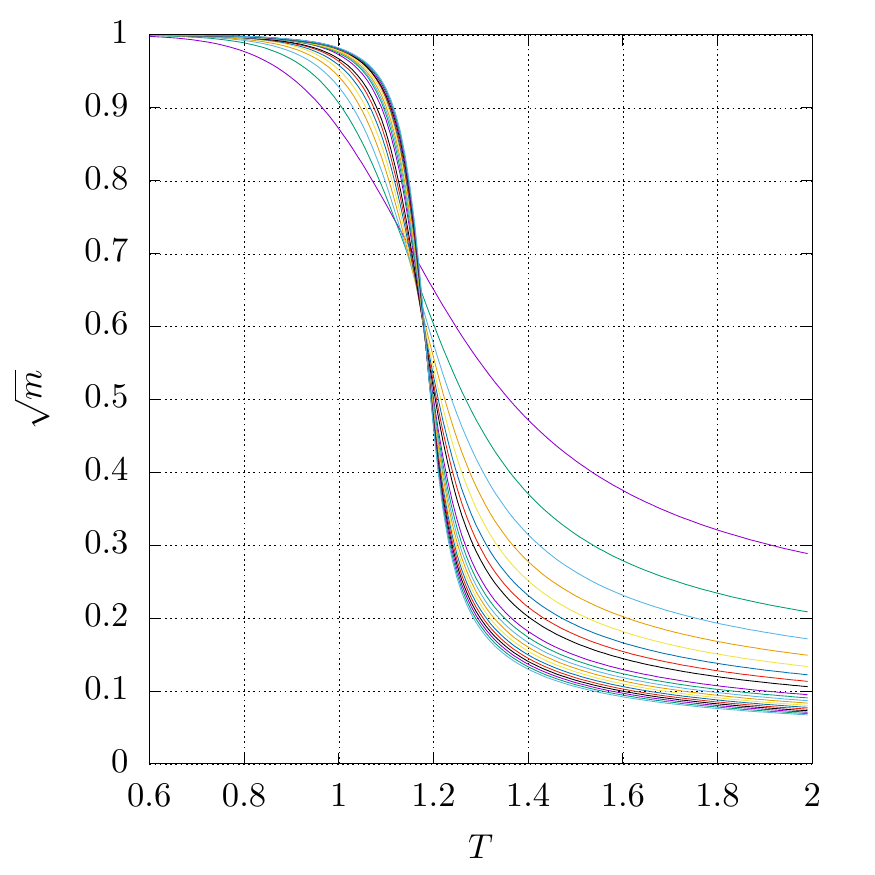}
\caption{Magnetisation per spin for $h=0$ in 2d for lengths up to $N=1000$.}\label{magh02d}
\end{figure}

\begin{figure}[H]
\includegraphics[width=8cm]{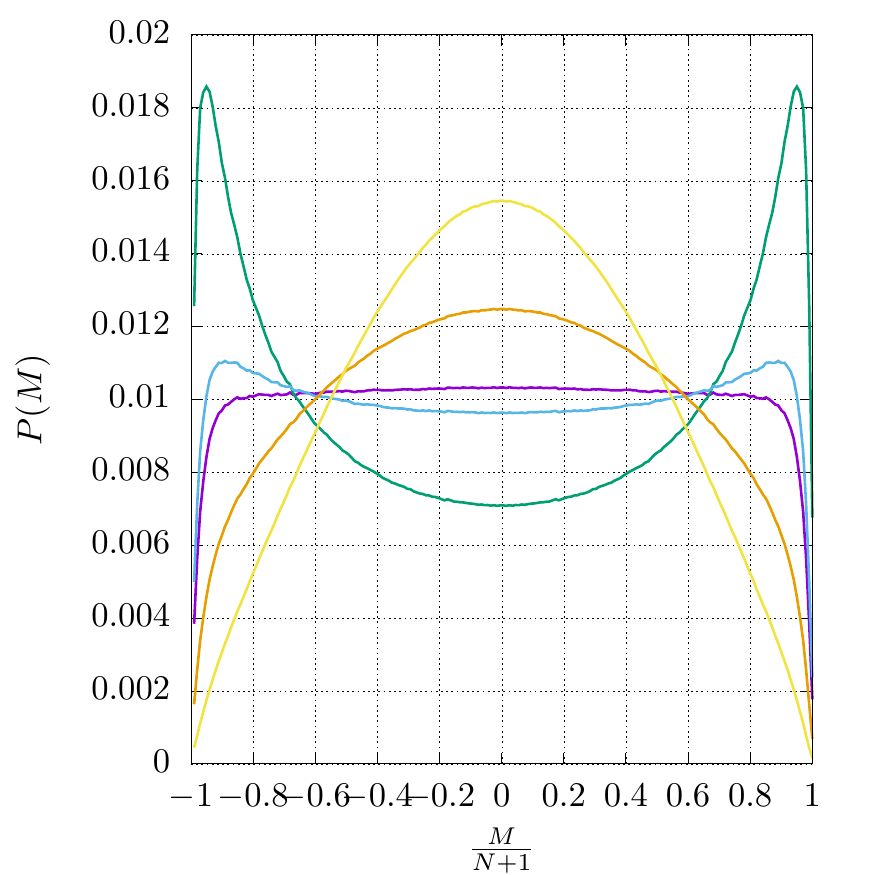}
\caption{$P(M)$, the probability distribution for 5 different values of $\beta$. $\beta=0.8,0,82,0.835,0.84$ and $0.86$ for $N=200$.}\label{2dh0}
\end{figure}

In figures~\ref{magh02d}--\ref{2dh0} we present results for $h=0$ in two dimensions. This includes graphs for the magnetisation per spin,  $\langle R_g^2\rangle/N$, which is simply the inverse of the unnormalised density, the Binder cumulant $U_m$ and the probability distribution for different values of the magnetisation per spin at different values of $\beta$, including one close to the expected transition $\beta_c\approx 0.835$, or $T=1.20$. Unlike the three-dimensional case, the form of $U_m$ is consistent with a critical transition. The peaks in Figure~\ref{2dh0} show clearly the two phase region for $\beta>0.835$, and the one phase region for $\beta<0.835$. At the transition, the profile is flat, and looking at the probability distribution of the energy (not shown) there is no evidence of coexistence between low-temperature and high-temperature phases. We conclude that the transition at $h=0$ is continuous in 2d, unlike what was observed in three dimensions.



 \begin{figure}[H]
\includegraphics[width=8cm]{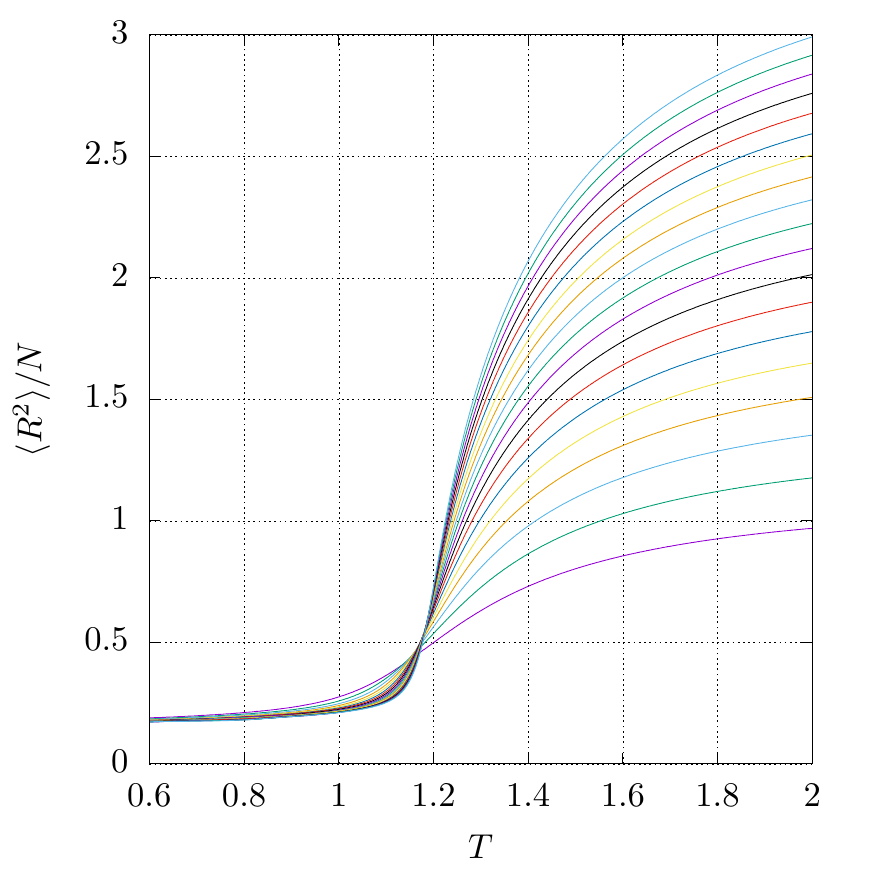}
\caption{$\langle R_g^2 \rangle/N$ for $h=0$ in 2d for lengths up to $N=1000$.}
\end{figure}
 \begin{figure}[H]
\includegraphics[width=8cm]{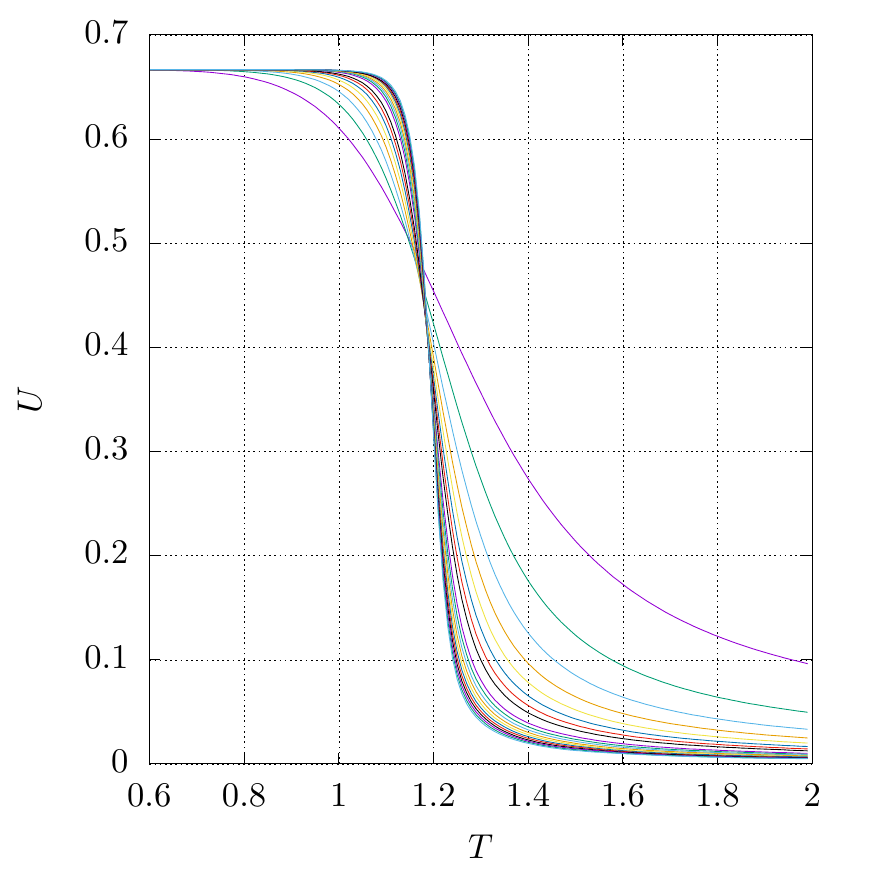}
\caption{Binder 4th order cumulant as a function of $T$ for sizes up to $N=1000$.}
\end{figure}



\begin{figure}[H]
\includegraphics[width=8cm]{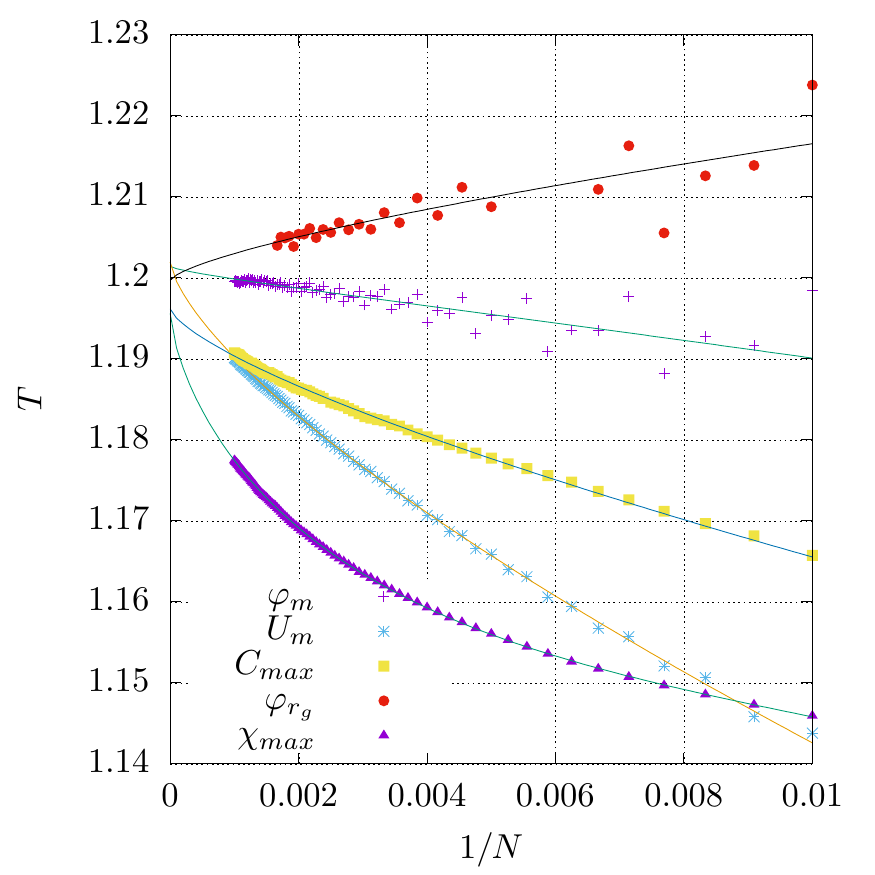}
\caption{Different estimates of the transition for $h=0$ using Binder's Cumulant $U_N$,  the scaling functions $\varphi_m$ and $\varphi_{R_g}$, as well as the locations of the peaks of $C$ and $\chi$. The lines are quadratic fits (or cubic in the case of $\chi$) in $N^{-\phi}$ with $\phi=0.689$.}\label{2dTc}
\end{figure}

In Figure~\ref{2dTc} we show different estimates of $T_c$ using the peaks of the specific heat, susceptibility, crossings of Binder's Cumulant and phenomenological RG using $\varphi_m$ and $\varphi_{R_g}$. 
Estimates of the transition temperature are calculated and shown in Figure~\ref{2dTc} using various methods: phenomenological renormalisation group, using the finite-size scaling forms of different quantities such as  $\varphi_m$ and $\varphi_{R_g}$; crossings of $U_m$ and the location of the peaks of $C$ and the magnetic susceptibility $\chi$, defined through
\begin{equation}
\chi=\frac{1}{TN}\left(\langle M^2\rangle-\langle |M| \rangle^2\right).
\end{equation} 

We can gain an idea of the cross-over exponent by looking at the scaling of the imaginary part of the complex Boltzmann weight $\tau=\exp(-1/T)$,  calculated from the leading zero of the partition function $Z_N$ as before. This is expected to behave as $\tau_i\sim N^{-\phi}$, which leads to an estimate of $\phi\approx 0.689$. 
This estimate is found by fitting the $\ln \tau_i$ vs $\ln N$ for large $N$ ($\ln N>6$). Varying the slope of the fitted line around this value gives a range of acceptable values of around $\phi=0.7\pm0.03$. The best fits in this section were found using $\phi=0.689$ for the finite sizes we have here. The finite-sized estimates are fitted by quadratics in $tN^\phi$, except the estimates from the peaks of $\chi_N$, which were fitted with a cubic. The corresponding lines are also shown in Figure~\ref{2dTc}. Using $\phi=0.689$, the fits give extrapolated temperatures leading to $T_c=1.199\pm 0.003$.

\begin{figure}[H]
\includegraphics[width=8cm]{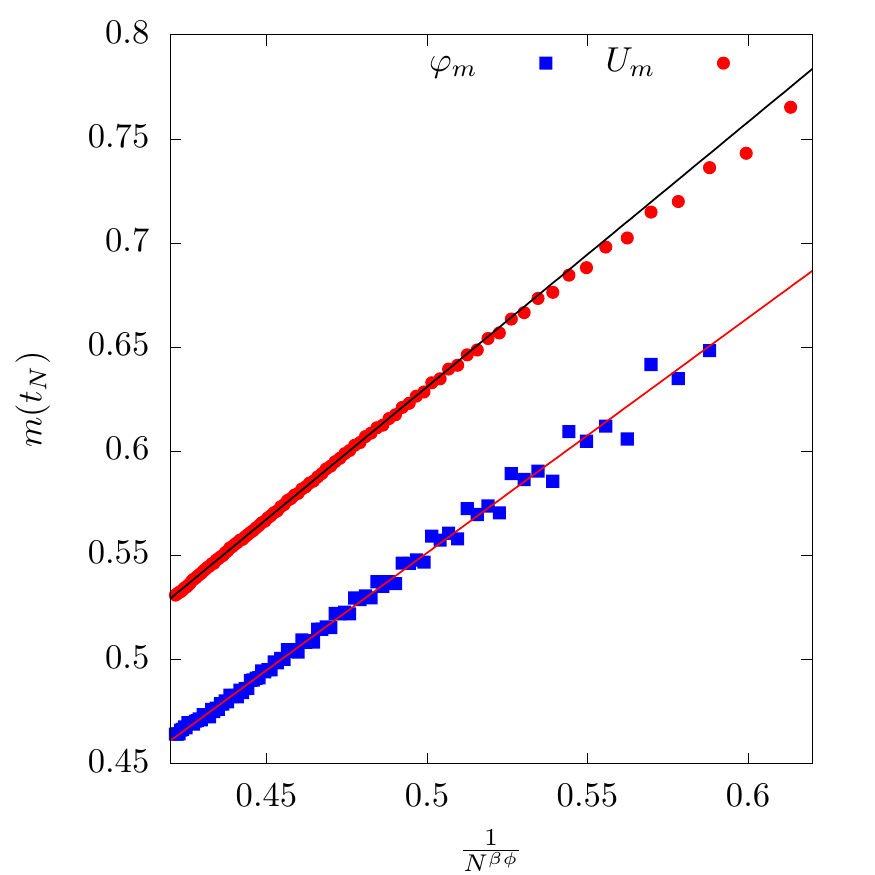}
\caption{Magnetisation calculated at the finite-size estimate of the critical temperature from the crossings of $\varphi_m$ and the Binder Cumulant plotted as a function of $N^{-\beta\phi}$ with \mbox{$\beta\phi=1/8$}.}\label{2dmag}
\end{figure}
\begin{figure}[H]
\includegraphics[width=8cm]{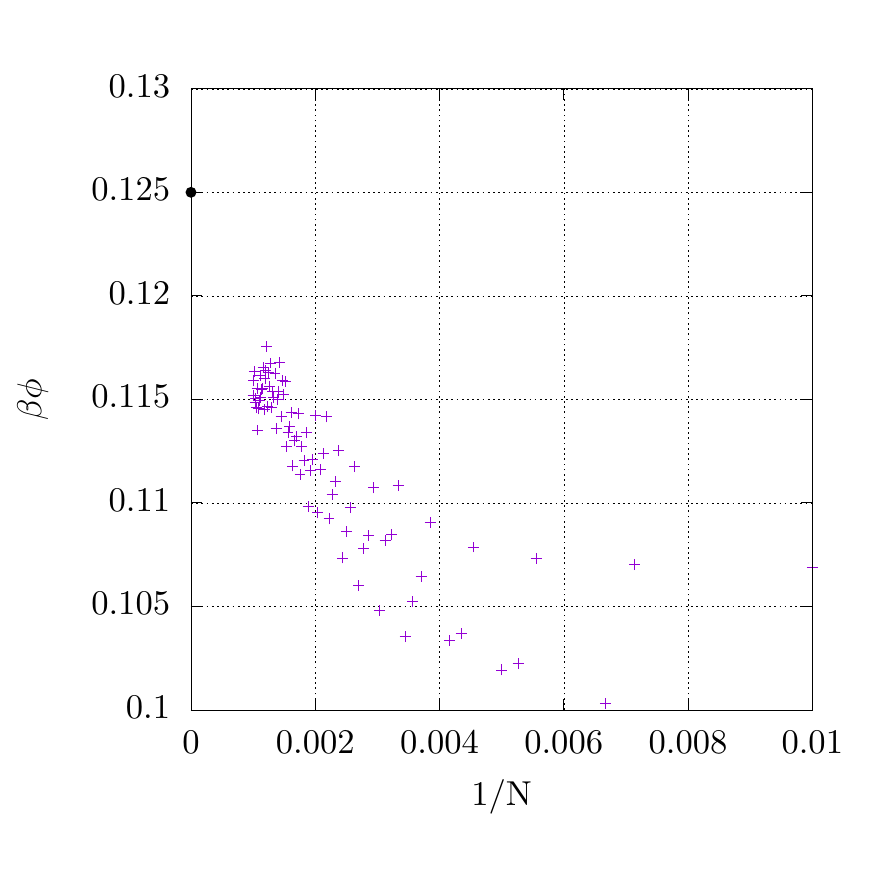}
\caption{Estimates of $\beta\phi$ from crossings of $\varphi_m$ plotted as a function of $1/N$.}\label{PRGmag}
\end{figure}

In Figure~\ref{2dmag} shows the magnetisation calculated at the estimates of $T_c$ for both crossings of $\varphi_m$ and $U_m$. The solid lines are plotted using a form $m_N=AN^{-1/8}$, and the fit to the curve is very good. Figure~\ref{PRGmag} shows the direct calculation of the $\beta\phi$ estimates, plotted against $1/N$, with a point at $\beta=1/8$ shown. It is curious that $\beta\phi$ should be so close to the exact value of $\beta=1/8$ for the two-dimensional Ising model.

\begin{figure}[H]
\includegraphics[width=6cm]{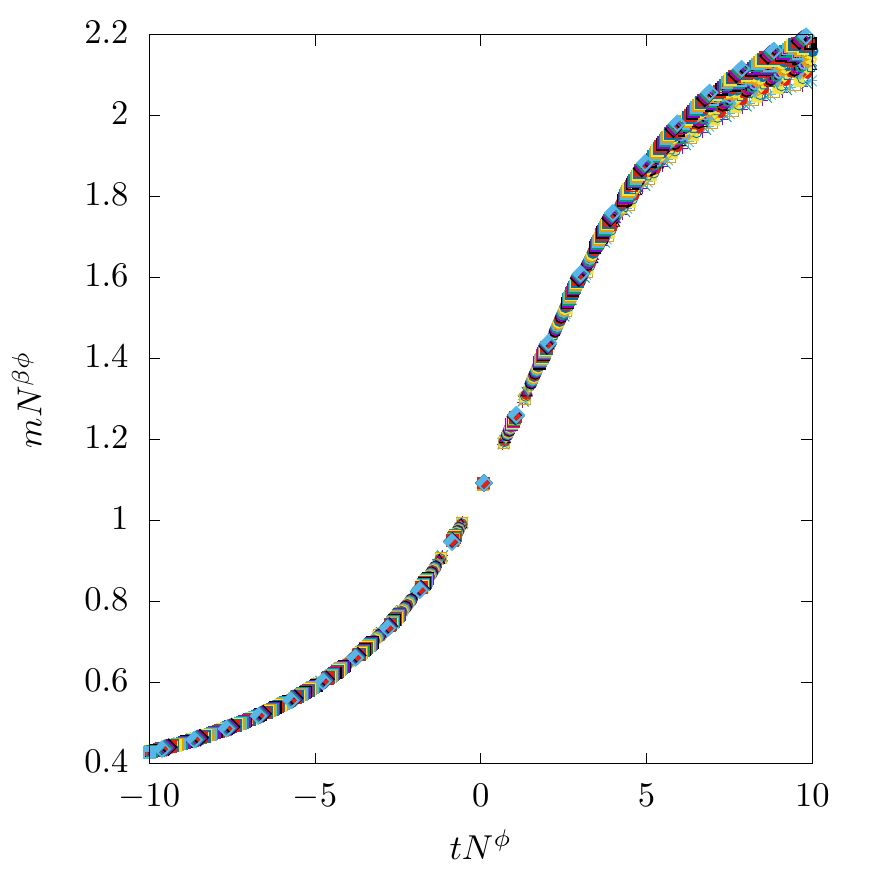}
\caption{Scaled magnetisation $mN^{\beta\phi}$ with $\beta\phi=1/8$, $\phi=0.689$ and $T_c=1.201$ with $N$ from 500 to 1000 in steps of 10.}\label{chi}
\end{figure}\label{magDC}

We further check on the exponents by looking at the data collapse of the magnetisation curves, shown in Figure~\ref{magDC}. We obtain good collapse of the data using $\beta\phi=1/8$, $\phi=0.689$ and $T_c=1.201$.

\begin{figure}[H]
\includegraphics[width=8cm]{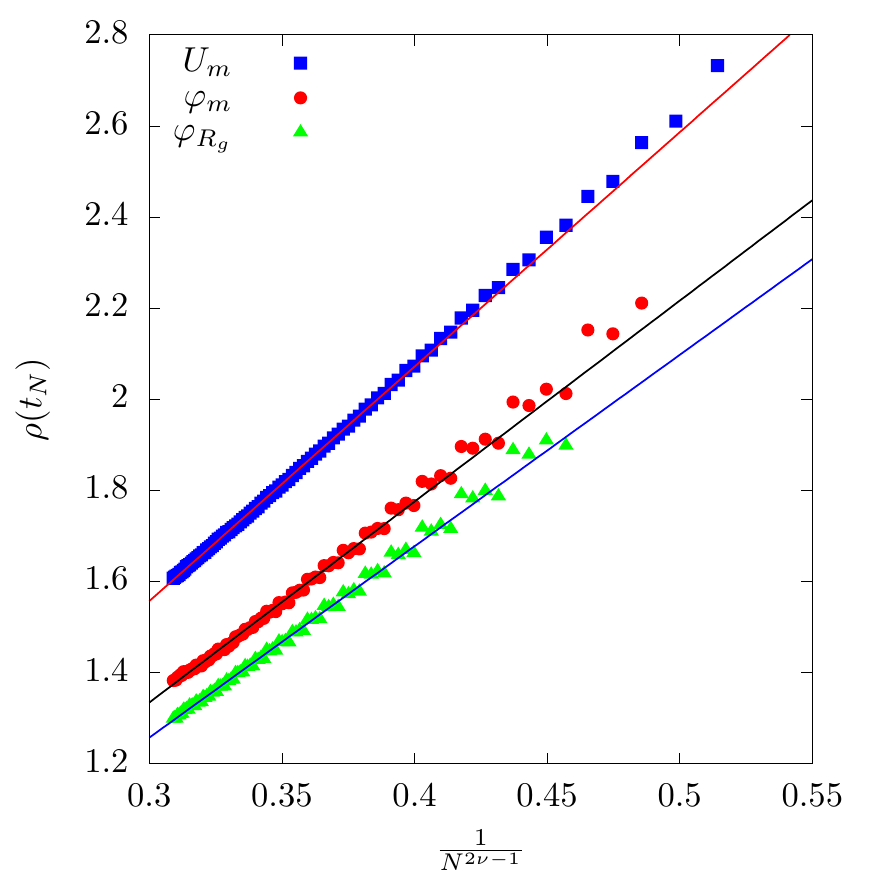}
\caption{the value of the un-normalised density $\rho_c=N/\langle R_g^2\rangle$ measured at the crossing value of $\varphi_{R_g}$ and $\varphi_m$ for $N, N/2$ and $N/2,N/4$ as well as at the crossings of $U_m$ for $N$ and $N/2$. The estimates are plotted as a function of  $N^{1-2\nu}=N^{-0.17}$, or $\nu=0.585$.}\label{2drho}
\end{figure}

\begin{figure}[H]
\includegraphics[width=8cm]{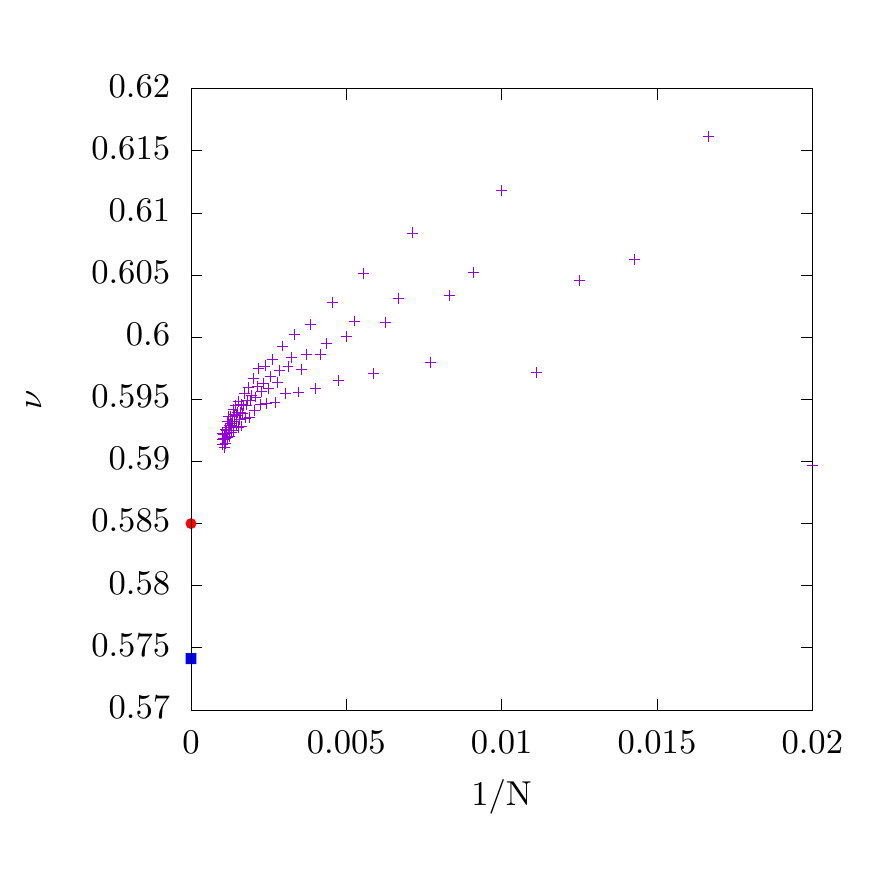}
\caption{Estimates of the exponent $\nu$ as a function of $1/N$, calculated from the crossings of $\varphi_{R_g}$. We compare with the estimate found above $\nu=0.585$ and $\nu=4/7$ corresponding to the standard collapse transition.}\label{PRGrho}
\end{figure}

Figure~\ref{2drho} shows the unnormalised density calculated at the finite size estimates of $T_c$ plotted against $1/N$. The curves are fitted to calculate $\nu$ from the expected behaviour $\rho_{c,N}=AN^{1-2\nu}$. This is done by a best fit of the $\log-\log$ curve of $\rho_c$ against $N$ for large enough $N$. The fit is seen to be good, giving $\nu=0.585\pm0.01$, slightly higher than the usual collapse transition value of $\nu=4/7\approx 0.5714\cdots$. Figure~\ref{PRGrho} shows the direct calculation of $\nu_N$ from the crossings of $\varphi_{R_g}$, which are consistent with the higher value of $\nu=0.585$, but could also include $\nu=4/7$.





\begin{figure}[H]
\includegraphics[width=6cm]{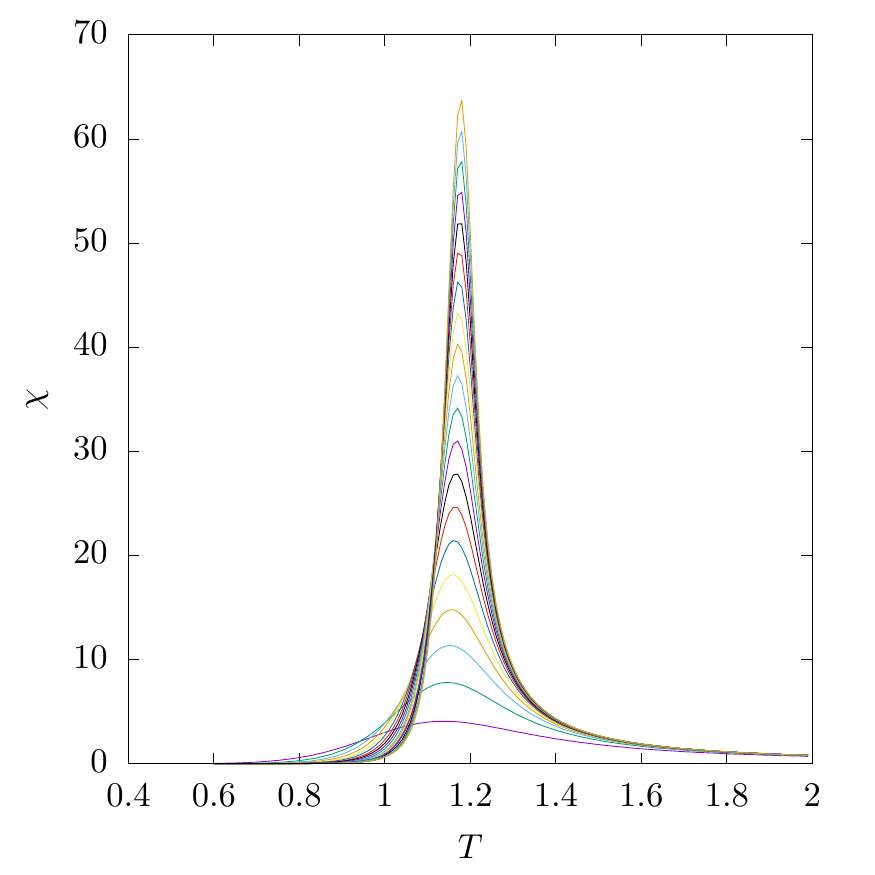}
\caption{Magnetic susceptibility $\chi$ plotted against $T$ for sizes from $50$ to $N=1000$ in steps of 50.}\label{chi}
\end{figure}

\begin{figure}[H]
\includegraphics[width=6cm]{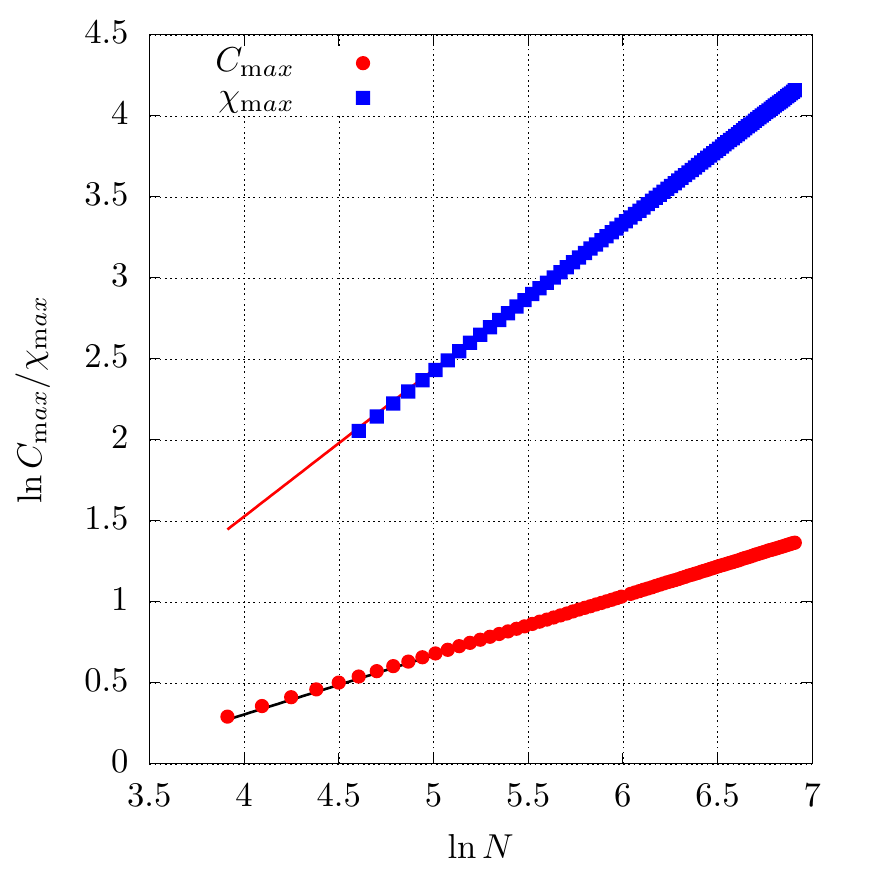}
\caption{ $\ln(\chi_{\rm max})$ and $\ln C_{\rm max}$ plotted vs $\ln(N)$. The slope of the line is fitted to a line of slope $\gamma\phi=0.905$ for $\chi_{\rm max}$ and $\alpha\phi=0.365$ for $C_{\rm max}$.}\label{logchi2d}
\end{figure}

\begin{figure}[H]
\includegraphics[width=6cm]{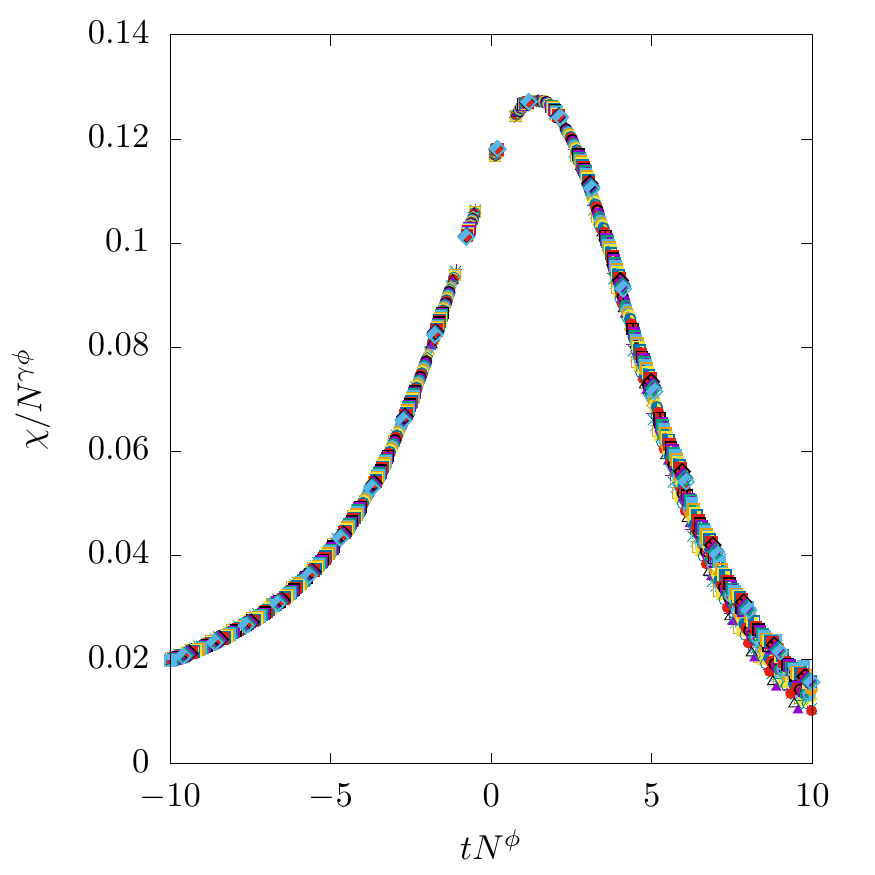}
\caption{Scaled susceptibility $\chi/N^{\gamma\phi}$ vs $tN^\phi$ with $T_c=1.192$, $\gamma\phi=0.9$ and $\phi=0.689$ with $N$ from 500 to 1000 in steps of 10.}
\end{figure}

\begin{figure}[H]
\includegraphics[width=8cm]{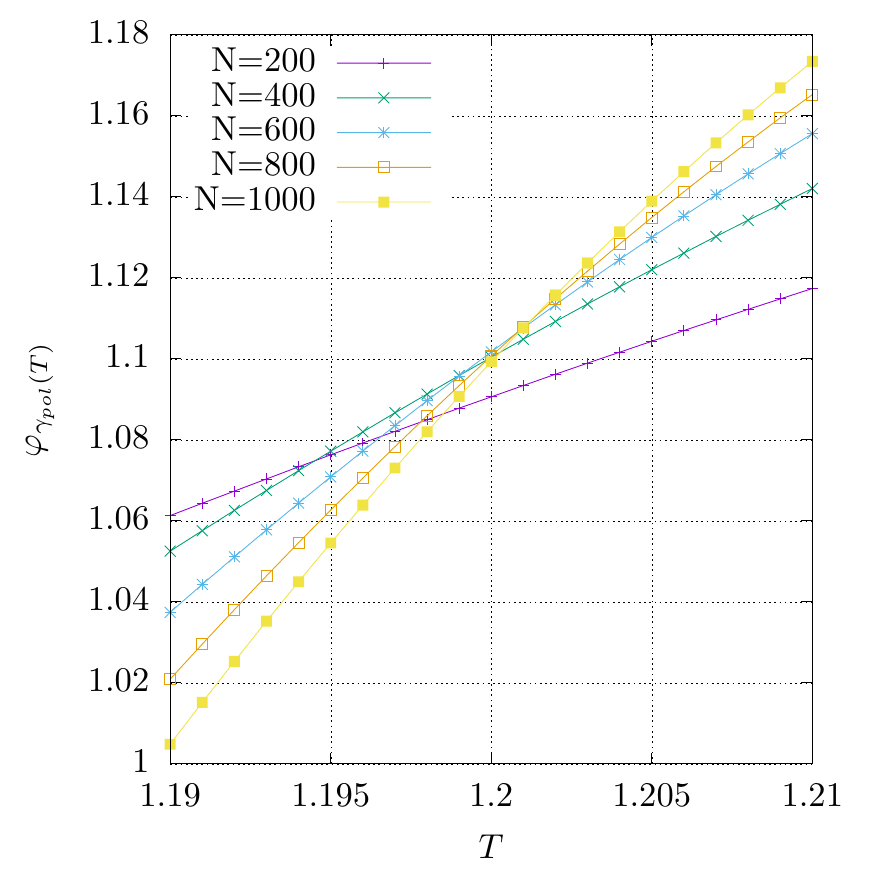}
\caption{$\varphi_{\gamma_{\rm pol}}$ for sizes $N=200, 400, 600, 800$ and $1000$}\label{varphigammapol}
\end{figure}

\begin{figure}[H]
\includegraphics[width=8cm]{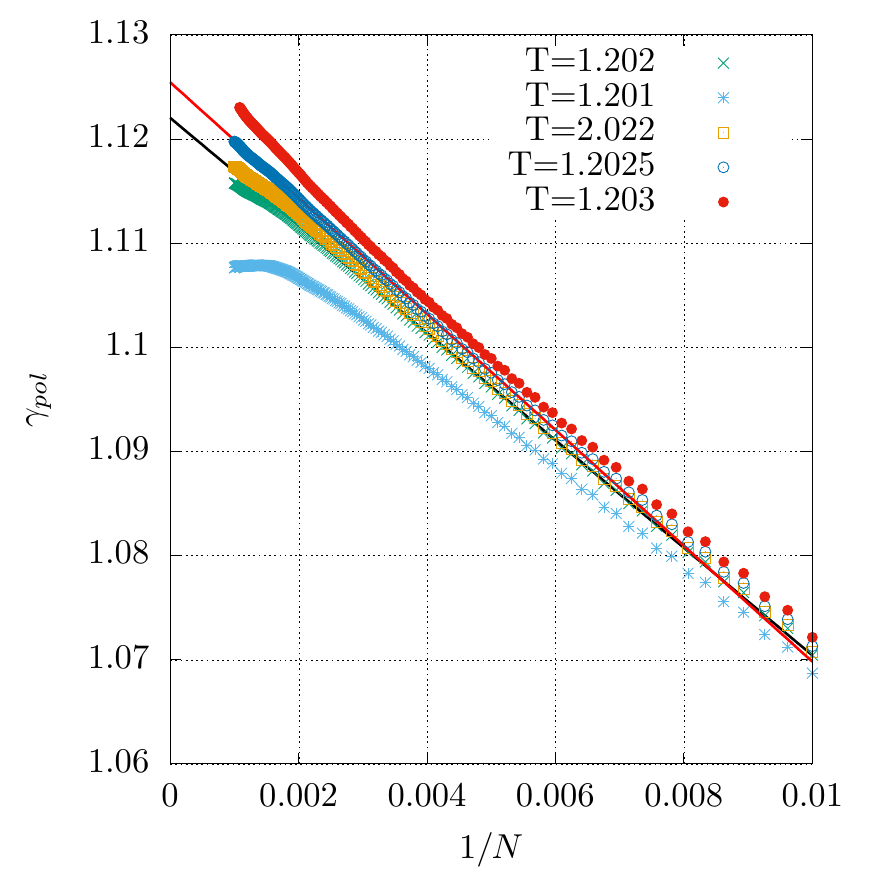}
\caption{plots of estimates of $\gamma_{\rm pol}$ as a function of $1/N$ for different temperatures around the critical temperature estimated from $\varphi_{\gamma_{\rm pol}}$ }\label{critregion}
\end{figure}

In Figure~\ref{chi} we plot the magnetic susceptibility, which shows peaks growing and approaching the critical temperature $T_c\approx 1.2$. In Figure~\ref{logchi2d} we plot the log of $\chi_{\rm max}$ against log $N$, and fit to a straight line to get an estimate of $\gamma\phi$. we find $\gamma\phi=0.905$ from the fit, but this could drop a little. We fitted the portion of the line from $\ln N=6$, and the asymptotic slope might be a little steeper. Taking the expression of $\phi=0.689$ would give $\gamma=1.31$. Using the higher and lower limits of $\phi=0.69\pm0.02$ leads to $\gamma=1.31\pm0.03$. In the same figure, we plot $\ln C_{\rm max}$ against $\ln N$. Likewise, fitting with a straight line gives $\alpha\phi=0.365$. Using $\alpha=2-1/\phi$, we find $\phi=0.683$, which confirms the value of $\phi$ found from the complex partition function zeros. Substituting, gives $\alpha= 0.54\pm 0.01$.

Since we have the full set of effective coefficients $C^{\rm eff}_{N,n_i}$ up to $N=1000$, we can look at the cut-off grand-canonical partition functions
\begin{equation}\label{gcpf}
{\cal Z}_N=\sum_{l=0}^{N} K^l Z_l.
\end{equation}
$N$ here is the maximum length taken into account in the cut-off partition function. The canonical partition function is expected to scale like 
\begin{equation} Z_N\sim \mu^N N^{\gamma_{\rm pol}-1},\end{equation}
where $\gamma_{\rm pol}$ is the polymer entropic exponent, not the exponent related to the magnetic susceptibility, defined above. The naming convention comes from the identification of the walk configurations in ${\cal Z}$, the grand-canonical partition function, as the graphs coming from the high-temperature expansion of the susceptibility of the $O(n=0)$ spin model\cite{Blote1989}. 

Substituting into Equation~(\ref{gcpf}) gives the grand-canonical scaling behaviour
\begin{equation}
{\cal Z}\sim (K_c-K)^{-\gamma}.
\end{equation}
Using the scaling of average length with $K$, which is now limited by the cut-off in length, gives the finite-size scaling relation:
\begin{equation}
{\cal Z}\sim N^{-\gamma}.
\end{equation}
 We can define a phenomenological scaling function
 \begin{equation}
 \varphi_{\gamma_{\rm pol}}=\frac{\ln\left(\frac{{\cal Z}_N}{{\cal Z}_{N/2}}\right)}{\ln 2}.
 \end{equation}
 The crossings of these functions will give estimates of both $\gamma_{\rm pol}$ and $K_c$.
 Figure~\ref{varphigammapol} shows, for each value of $T$ shown in the critical region, the estimated $\gamma(T)$ calculated from the crossings of $\varphi_{\gamma_{\rm pol}}$ for $N/(N/2)$ with $(N/2)/(N/4)$ plotted for different values of $N$ in the critical region. Intersections of these estimates will further pick out estimates for $T_c^N$. Looking at the intersections and extrapolating leads to an estimate of $T_c\approx 1.202$. This can be examined further by fixing the expected critical temperature, and looking at the estimate of $\gamma_{\rm pol}$ as a function of length. If we are at the critical temperature, the exponent value is expected to be linear in $N$ for large $N$, from the linearity of $k=(K_c-K)/N$ in $N$. If we are off the critical temperature, but close, the smaller sizes will be in the critical region, but as the length increases, this region becomes smaller, and the walk is subject to cross-over effects, pulling the estimate off linearity. This is shown in Figure~\ref{critregion}, where the best fit to a linear line up to $N=1000$ is $T_c=1.2025$, which is a little higher than found with a canonical analysis above, but still within the expected error bars. This gives $\gamma_{\rm pol}=1.1255\pm 0.003$, which is a little smaller than the exact value of $\gamma_{\rm pol}=8/7\approx 1.143$ at the standard collapse transition, but fairly close to it.

\section{Discussion}

In this paper we have revisited the three-dimensional magnetic polymer, realised by a chain of Ising spins along the backbone of a fluctuating walk, with the Ising energy as the sole interaction energy. We have produced improved numerical results, which clearly show the nature of the first-order simultaneous collapse and magnetic transition. Both appear to be first order, but the average radius of gyration still scales as $R_g\sim N^{1/2}$, rather then $R_g\sim N^{1/3}$ as expected for a dense walk. 
The mean-field theory of Garel {\it et al.}\cite{Garel1999} predicts a first-order transition for small $h$, becoming the standard collapse transition for larger $h$. They claim that the first-order transition exists for finite, non-zero, magnetic field, based on the scaling of the specific heat. We find similar results, but when we look at the probability distribution of both $M$ and $n_i$, even with $h/T=0.1$, there is only a marginal sign of the possible existence of a finite-size first-order transition. The question of a finite $h$ first-order transition is, in our mind, still open. 

This MFT does not restrict the model to a single walk, and equally applies to a ``gas" of walks, and it might be interesting to see how removing the restriction of a single walk affects the  behaviour in both two and three dimensions. 

When the three-dimensional model is changed to introduce more fluctuations, such as the fluctuating bond model studied by Luo\cite{Luo2006}, the first-order gives way to a second-order transition. If we reexamine their scaling relations, and realise that they should have used $\phi$ and not $1/\nu$ in their scaling relations, we reinterpret their results as being $\phi=1$, $\beta\approx 1/3$, to be compared with the three dimensional Ising value of $\beta\approx 0.326$. In two dimensions, fluctuations are also more important, and again, we see a second-order transition. Interestingly, whilst we find a value of $\phi\approx 0.69\neq 1$, the magnetic transition is characterised by the an exponent $\beta\phi=1/8$, to be compared with $\beta=1/8$ for the usual transition. This warrants further investigation.   

In two dimensions, we could find no evidence that there is a magnetic transition for $h\neq 0$, and it would seem likely that for $h\neq 0$ the transition is the usual collapse transition.  The geometric and entropic exponents found for the walk, $\nu$ and $\gamma_{\rm pol}$ are close to, but not the same as, the usual $\theta$-point exponent values.
This could be due to finite-size effects.

To summarise the results presented:

\begin{widetext}

\begin{center}
\begin{tabular}{|c|c|c|c|c|c|c|}\hline
 & $T_c$ & $\alpha$ & $\phi$ &$\gamma$ & $\nu$ & $\gamma_{\rm pol}$\\\hline
2d & $1.199\pm 0.003$ & $0.54\pm0.01$ & $0.69\pm 0.02$ & $1.31\pm0.03$ & $0.585\pm 0.01$ & $1.1255\pm 0.0003$\\\hline
3d & $1.90\pm 0.02$  & $1.06\pm x$ & 1 &  -- & 1/2  & --\\\hline
\end{tabular}
\end{center}
\end{widetext}

In the model presented here, the magnetic and collapse transitions occur together. It is simple to imagine a model where these two transitions, with different order parameters, might occur separately, which might give interesting additional critical behaviours. 

Recently a dynamic HP model has been studied in two dimensions\cite{Faizullina2021}. This model is similar in that both the H/P state of each monomer and the conformation are dynamic variables. The two main differences with the work here is that there is not an interaction along the chain, only between non-consecutive nearest-neighbour monomers of type P. They found that the collapse transition here was in the standard $\theta$ collapse universality class. 
 
 \acknowledgements
 
 The authors would like to thank Martin Weigel for useful comments.

%

\end{document}